\documentclass[journal]{vgtc}                

\usepackage{microtype}                 
\PassOptionsToPackage{warn}{textcomp}  
\usepackage{textcomp}                  
\usepackage{mathptmx}                  
\usepackage{times}                     
\usepackage{cite}

\usepackage{caption}
\usepackage{morefloats}
\usepackage[tight,normalsize,sf,SF]{subfigure}

\usepackage{amsmath,amscd}
\usepackage{amsfonts}
\usepackage[ruled,vlined]{algorithm2e}
\usepackage{epsfig}
\usepackage{enumerate}
\usepackage{multirow}
\usepackage{array}

\onlineid{0}

\vgtccategory{Research}

\title{MAC Rate Adaptation and Cross Layer Behavior for Vehicular WiFi Access: An Experimental Study}

\author{Zafar Ayyub Qazi, Saad Nadeem, and Zartash Afzal Uzmi}

\abstract{Vehicular WiFi is different from conventional WiFi access. Firstly, as the connections arise opportunistically, they are short lived and intermittent. Secondly, at vehicular speeds channel conditions change rapidly. Under these conditions, the MAC bit rate adaptation plays a critical role to ensure that devices can maximize throughput performance. The goal of this work is to gain a deeper understanding of current MAC bit rate adaptation algorithms and their interaction with higher layer protocols under vehicular settings. Towards this end, we evaluate the performance of four MAC bit rate adaptation algorithms; SampleRate, AMRR, Minstrel, and RRAA, and their interactions with transport layer protocols such as TCP in real world vehicular settings. Our experimental study reveals that in general these algorithm perform poorly in vehicular settings, and can have an extremely adverse impact on TCP performance.

} 

\nocopyrightspace

\vgtcinsertpkg

\begin{document}


\firstsection{Introduction}

\maketitle

There are a growing number of users accessing the Internet while in motion, in particular in vehicles e.g. email, web, VOIP etc. As a result more and more mobile devices are supporting computing and wireless communication ``on the go''.  Such mobile devices typically use cellular infrastructure but an increasing number of these now also come equipped with WiFi like Apple's iPhone, Samsung's BlackJack, Nokia 600 series etc. WiFi access is appealing because it is cheaper and supports higher data rates, besides being widely available. However, vehicular WiFi access is different from conventional WiFi access. Firstly, as the connections arise opportunistically, they are short lived and intermittent. Secondly, at vehicular speeds channel conditions change rapidly. Therefore protocols that perform well in traditional WiFi settings would not necessarily perform well in these settings. Prior work \cite{bychkovsky,gass,hadler,ott} has studied the performance issues in vehicular WiFi settings. Observations on connection setup-related protocols, transport layer protocols and applications have been well documented. However the performance of MAC bit rate selection algorithms and their interaction with higher layers has not been studied in detail.

MAC bit rate selection algorithms are designed to exploit the multi-rate capability of 802.11 networks, by attempting to select the transmission rate, best suited to the channel conditions.  The current 802.11 specifications allow multiple transmission rates at the physical layer (PHY) that use different modulation and coding schemes, e.g. the 802.11b PHY supports four transmission rates (1-11 Mbps), the 802.11a PHY offers eight rates (6-54Mbps), and the 802.11g PHY supports twelve rates (1-54Mbps)\cite{wong}. When the channel conditions are changing rapidly, these algorithms become even more important. These algorithms are expected to adapt MAC bit rate to changing channel conditions, in order to maximize throughput performance.

In this work, we consider Internet access in vehicles, in particular, short-lived connections to roadside 802.11 access points that arise opportunistically as vehicles are in motion. We conduct real outdoor experiments, to investigate the performance of different rate adaptation algorithms, their interaction with higher layers and their impact on the overall connection performance. Specifically, we test four rate  adaptation algorithms namely RRAA \cite{wong}, SampleRate \cite{bicket}, AMRR\cite{lacage}, and Minstrel \cite{minstrel}, along with TCP bulk traffic and CBR traffic over UDP.  We also test with fixed MAC bit rates (using all 802.11g bit rates), to understand the performance with each fixed MAC bit rate. We report our observations from over 168 experimental runs, including insights into connection setup protocols. Below, we highlight some of the key insights from our experimental study.
\begin{itemize}
\item We observe that ARP timeouts and TCP connection timeouts can cause significant delay in initial connection setup. The main reason is the high initial MAC bit rate used by the rate adaption algorithms to transmit ARP messages and TCP SYN, SYN+ACK, ACK messages.

\item We observe OFDM and DSSS rate have different characteristics in terms of their transmission range and RSSI threshold for demodulation. DSSS rates like 11Mbps and 5.5Mbps provide steady throughput performance whereas OFDM rates like 54Mbps and 12Mbps provide high throughput but only for a short time interval, when the RSSI value is high. 

\item All the four rate adaptation algorithms used high initial MAC bit rates (e.g., 54 Mbps), often are too slow to adapt to changing channel conditions in vehicular settings (using either a bit rate that is too high or too low), do not take into account the different characteristics of DSSS and OFDM rates and can frequently cause TCP retransmissions.
\end{itemize}

\section{Related Work}
Existing work on data communication in vehicular networks can be broadly classified as focusing on either vehicle-to-vehicle or vehicle-to-infrastructure communication. In this paper, we focus on the latter by considering data communication between a vehicle and access points that are part of an 802.11 WiFi infrastructure, specifically in a non-urban environment such as along the highways or in a rural setting.
\newline
The Drive-Thru Internet project was one of the pioneer works that studied the feasibility of using Internet access in vehicles via communication with roadside 802.11 access points. Under this project, Ott and Kutscher were able to transfer a maximum of 8.8MB data using UDP traffic and 6MB data using TCP traffic at 80 km/h \cite{ott}. They conducted experiments using external antennas with off-the-shelf 802.11b hardware. They used different vehicular speeds to gauge the differences in total data transferred, over a connection range of approximately 600m. They observed that the total connection time was inversely proportionally to the vehicle speed. They also identified three phases during the connectivity period, namely the entry, production and exit phases. The entry and the exit phases, as they observed, incurred large packet losses (that consequently lead to large amount of retransmissions in TCP during these phases). In a later study, Ott and Kutscher were able to transfer 20-70MB data at 120km/h using external antennas with 802.11g hardware \cite{ott1}.
\newline
Gass et al. conducted a similar feasibility study with 802.11b in vehicle-to-infrastructure settings. They investigated the effects of three parameters: car speed (5mph, 35mph, 75mph), network traffic type (UDP bulk traffic, TCP bulk traffic and web traffic) and backhaul network performance (1Mbps bandwidth limit and 100ms latency each way) using standard laptops with no external antennas \cite{gass}. They conjectured that at high speeds, the entry and exit phases are of smaller duration (but are more lossy) and that the mobile client almost enters the production phase directly. They validated the observation of \cite{ott} in reference to the inverse proportionality of the connection time and the vehicle speed. Furthermore they also corroborated the hypothesis of \cite{ott} that the backhaul parameters can significantly reduce the total data transferred during a single pass in the TCP session.
\newline
Cottingham et al. followed \cite{gass} with a performance evaluation of 802.11a, restricting their study to realistic urban speeds (7km/h and 45km/h) and environments \cite{cottingham}. They explored the effects of using different CBRs (10Mbps and 30Mbps) with UDP. In effect, they observed significantly larger variations in throughput, when using CBR of 30Mbps. They attributed these variations to the rates selected and subsequently, the coding schemes used: 48 and 54Mbps bit rate with 64-QAM coding scheme in case of 30Mbps, and 12 and 18Mbps bit rate with QPSK coding scheme in case of 10Mbps. They elaborated that 64-QAM, being more sensitive to interference than QPSK, incurred more packet losses for a given amount of noise.
\newline
Being feasibility studies, \cite{ott,ott1,gass,cottingham} they did not study the impact of the rate adaptation algorithms running at the MAC layer. \cite{shankar,chen,vutukurus} took this impact into consideration in the outdoor vehicular settings. \cite{shankar} evaluated the performance of different rate adaptation algorithms in comparison to their own novel algorithm (CARS) in Pseudo-IBSS mode with real testbed experimentations using MadWiFi and simulations using NS. Similarly, \cite{chen} demonstrated the effectiveness of their new algorithm (RAM) in comparison to other default algorithms in MadWiFi using experimentation and simulations. \cite{vutukurus} used NS simulations to study the impact of their rate adaptation algorithm along with RRAA using TCP traffic in high mobility scenarios(in contrast to the UDP traffic used in \cite{shankar} and \cite{chen}).
\newline
Camp et. al implemented and experimentally evaluated the mechanisms deployed by state-of-the-art loss-triggered and SNR-triggered rate adaptation algorithms \cite{camp} using their WARP platform. They evaluated the performance of each mechanism by comparing its selected rate with the ideal rate, found through exhaustively tracing out the rate strategy that maximizes the throughput. They observed that, in high mobility scenario using vehicle-to-infrastructure communication, sequential rate stepping of the loss-triggered mechanisms could not track rapid changes in the environment, whereas SNR-based algorithms were able to accurately adapt to changing conditions.
\newline
Following \cite{camp}, Hadaller et al. studied the behavior of a modified SampleRate in a rural highway setting using TCP traffic \cite{hadler}. They claimed that their rural highway settings had repeatable channel conditions and that environmental awareness on part of the protocols could help in selecting better operating parameters to work in these settings. They drew useful inferences from their experimentation, regarding the connection setup delays, application initialization delays, and poor rate selection by the rate adaptation algorithm used. \cite{bychkovsky}, though focusing on the urban settings, similarly identified connection setup delays and application initialization delays as major factors in restricting the optimal utilization of useful connection period, in which a greater amount of data could be transferred.
\newline
We distinguish our work from the previous studies by comprehensively studying the interaction of different rate adaptation algorithms \cite{wong,bicket,minstrel,lacage} with higher layers in vehicle-to-infrastructure settings. In effect, we consider both UDP as well as TCP traffic during our experimentation. We also study the effects of varying the CBR in reference to UDP. We selected Sample Rate, AMRR, RRAA and Minstrel because they are representative of the auto-rate schemes that utilize the statistics-based approach; these algorithms use statistics collected in a particular time period to select an appropriate MAC bit rate. We were not able to implement any SNR-based algorithms because of the modifications required in the 802.11 standard.

\section{Experimental Setup}
\begin{figure}[t]
\centering
\epsfig{file=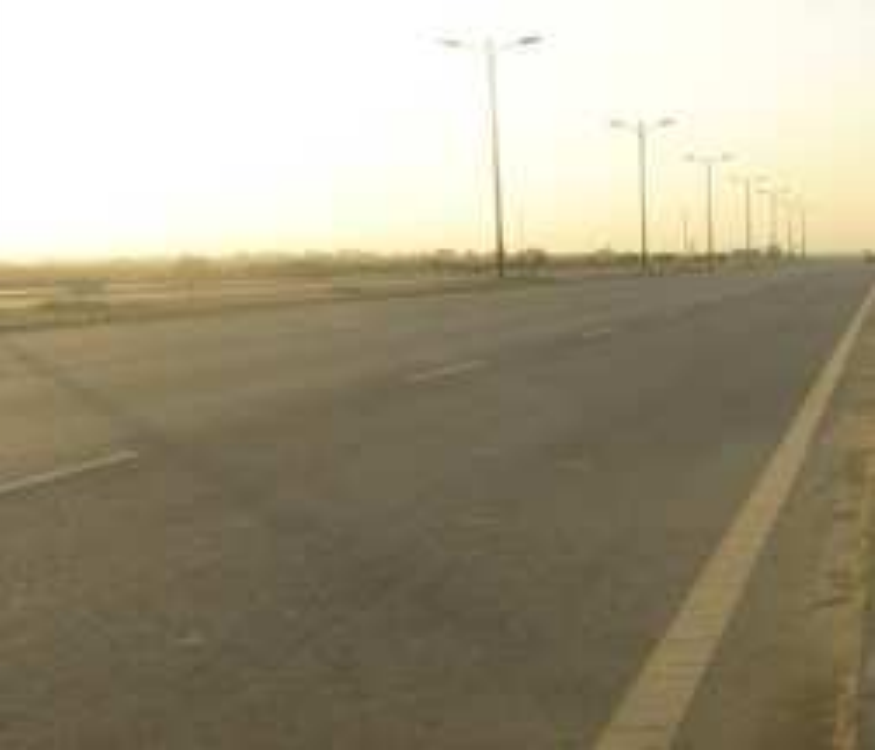, height=2in, width=3.3in}
\caption{The location where the experiments were conducted }
\label{fig:location}
\end{figure}

The focus of our experiments was to investigate the performance of rate adaptation protocols, their interaction with higher layers and their impact on the overall connection. We have conducted our experiments in a rural highway setting (DHA Phase VI, Lahore), namely on a straight flat road, with equally spaced street lamps on one side and plain field on the other(see Figure \ref{fig:location}). The road traffic was negligible. We have conducted several runs for fixed parameters to validate our observations and we perform in total over 168 experimental runs. The experiments were conducted across 15 days with three people.

\subsection{Experimental Parameters}
All our experiments were conducted at a speed of 50km/h. The affects of changing speed have been well studied in [4]. More specifically, [4] concluded that higher speeds do not introduce higher frame loss rates . We choose 50km/h instead of normal highway speeds because increasing the speed meant smaller connection duration and hence less connection time to analyze protocol behavior. The cards we used had transmission range of approximately 350m.
\newline
The four algorithms that we have used are implemented in Madwifi \cite{madwifi} drivers . We tested these algorithms with CBR traffic (10Mbps and 30Mbps with UDP) and TCP bulk traffic (simulating FTP file transfers). In our experiments we used downstream TCP and UDP traffic, we leave the consideration of upstream traffic to future work. The vehicle will likely initiate the TCP connection in practice, in our experiments, the TCP sender (the AP) had to initiate the connection due to a limitation of our traffic generation software.  We used statically configured IP addresses, as DHCP is well-known to behave poorly in this environment [3]. We have pre-configured the client with our AP ESSID.
\newline
We used the default transmit powers of 18dbm for both the access point and client. We switch off mechanisms that are proprietary to Atheros cards including Fast frames, Super A/G adaptive Radio, Atheros Bursting, Turbo and link layer compression. This configuration allows us to investigate the performance of 802.11 standard independent of card specific mechanisms and to see more clearly the affect of using different protocols and parameters on the overall performance. We also disable antenna diversity which is reported to have significant affect on the performance of 802.11 cards (the effects of using antenna diversity have been reported \cite{Giustiniano}) and is enabled in our cards by default.

\subsection{Hardware and Software}
\begin{figure}
\centering
\epsfig{file=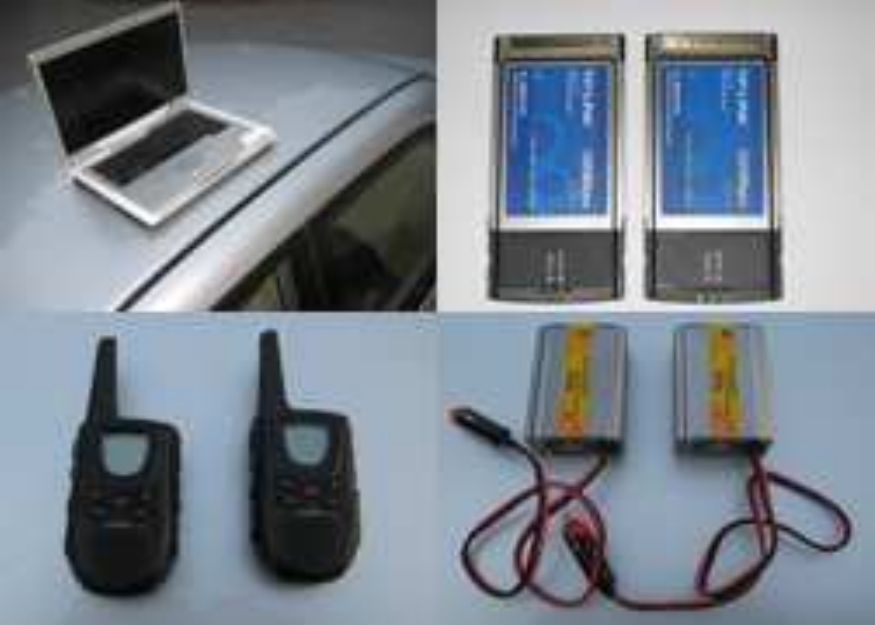 , height=2in, width=3.3in}
\caption{Equipment}
\label{fig:hardware}
\end{figure}

Table \ref{tab:hardware} summarizes the hardware and software used for our experimentation (also see Figure \ref{fig:hardware}). We had had two laptops, one configured as an AP and the other as client. Data was captured by putting the Atheros card in monitor mode and using tcpdump version 3.9.4 to capture all frames, including extra MAC layer information from the card in the radiotap header, such as the MAC bit rate and measured RSSI for each frame. We used wireshark-1.0.8 and libtrace 3.0.6 to analyze the tcpdump trace files. We used iperf-2.0.4 \cite{iperf} to send UDP data and bulk TCP data from the access point to the client.
\begin{table}[t]
\centering
\caption{Hardware, Software and Experimental Parameters}
\label{tab:hardware}
\begin{tabular}{|l | l|}
\hline
\textbf{Parameter} & \textbf{Value }\\
\hline
\hline
Client Laptop & Acer Travelmate 5710 with \\
 & 1.6GHz processor \\
  & and 1GB RAM \\ \hline
Access Point & Dell Inspiron 6000 with
\\ Laptop & 1.6GHz processor \\
& and 512MB RAM \\ \hline
Operating System & Linux
\\ & (Cent OS 5.2 \\
& kernel 2.6.18-53.e15)\\ \hline
Wireless Card & TP-Link TL-WN610G with \\
& TL-WN610G with \\
& Atheros 5212 chipset\\ \hline
Driver & Madwifi 0.9.4\\\hline
Traffic generating  & Iperf-2.0.4\\
& software \\ \hline
Direction of traffic & Downstream\\ \hline
Standard & 802.11g with \\
& all twelve rates\\ \hline
Active Analysis & Tcpdump 3.9.4+\\
& Libpcap 0.9.4\\ \hline
Passive Analysis & Libtrace 3.0.6\\
& wireshark-1.0.8\\ \hline
Frequency & 2.412GHz\\ \hline
Transmit Power & 18dbm for
 both AP and Client \\ \hline
Packet size & 1Kbytes\\ \hline
CBR & 10Mbps and 30Mbps\\ \hline
Location & Phase VI, DHA\\ \hline
Speed & 50km/h \\ \hline
\end{tabular}
\end{table}

\subsection{Experimental Procedure}
\begin{figure}[t]
\centering
\epsfig{file=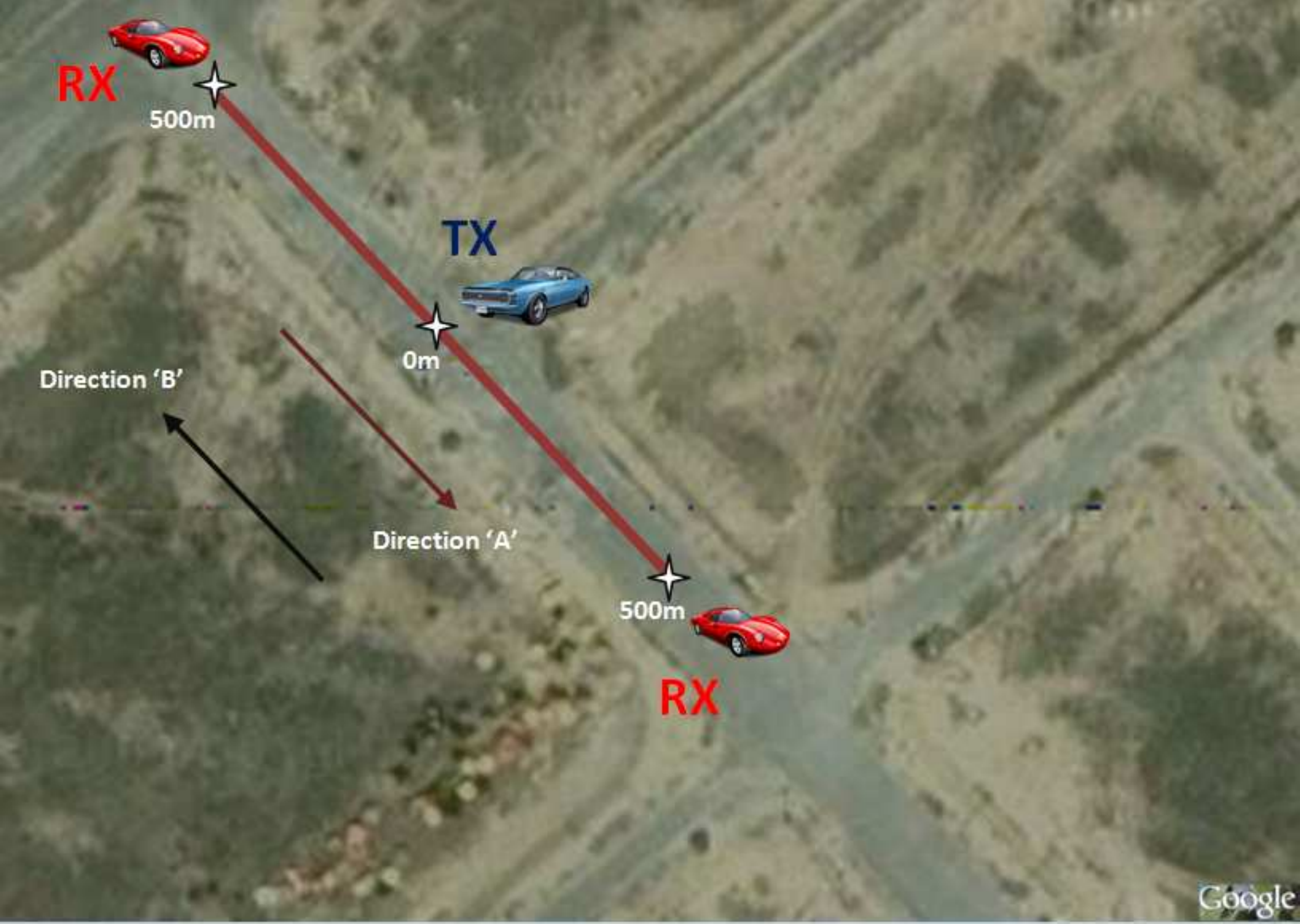 , height=2in, width=3.3in}
\caption{The Map of our settings}
\label{fig_map}
\end{figure}

Our experiments were conducted as follows. We mark the road 500m (well out of AP range) on either side of the AP. Each experiment begins with the client laptop on the lap of the passenger in the vehicle, well beyond the 500m distance. At this point the logging scripts and sniffers on the vehicle and access point are started. As the car hits the first 500m mark the driver  maintains a constant speed of 50km/h and simultaneously an enter key is pressed on both client and AP causing a timestamp to be recorded. The markers are communicated through the walkie-talkie. When the client comes into the range of the access point it automatically associates using the AP ESSID and begins sending traffic. In the case of TCP at the start of an experiment, the vehicular client runs iperf in listener mode, waiting for a connection from the iperf sender. Once the client enters range, it performs a standard MAC association with the access point. Using a shell script, the access point detects a newly associated client and launches the iperf sender, which initiates a bulk TCP connection to the statically configured client IP. When the vehicle reaches the access point, again a time stamp is recorded at both the client and access point. Finally when the car passes the end 500m mark, final timestamp is recorded. The timestamps provided the client's position relative to an access point (see Figure \ref{fig_map}).

\section{Analysis}
\begin{figure}[t]
\centering
\epsfig{file=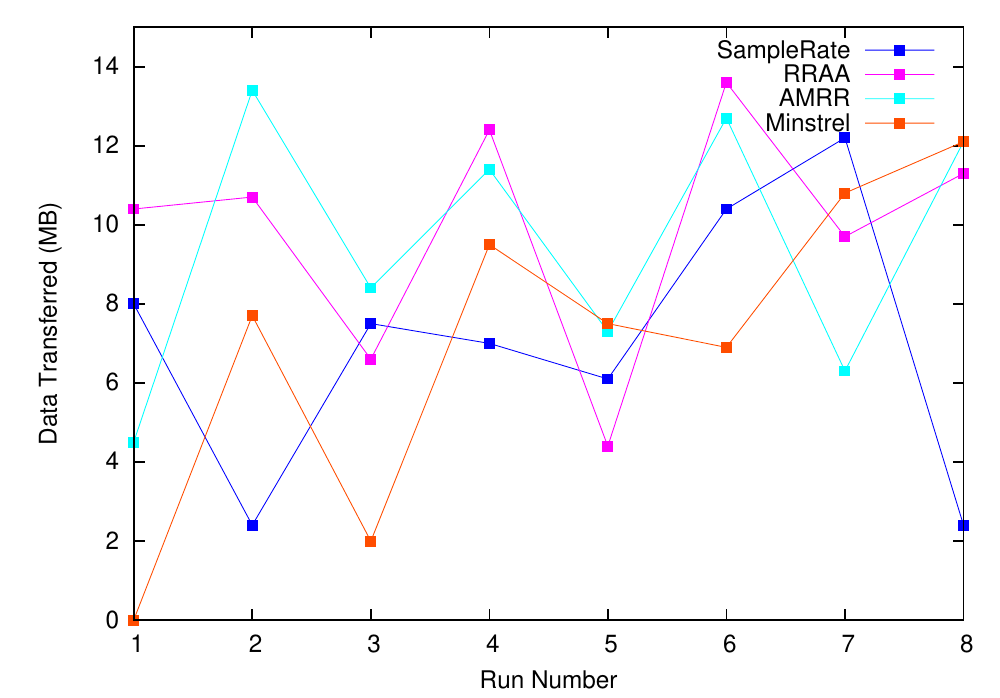 , height=2in, width=3.3in}
\caption{The variation in the amount of data transferred for RRAA, SampleRate, AMRR and Minstrel with CBR traffic of 30Mbps over UDP }
\label{fig:var_dt}
\end{figure}

\begin{figure}[t]
\centering
\epsfig{file=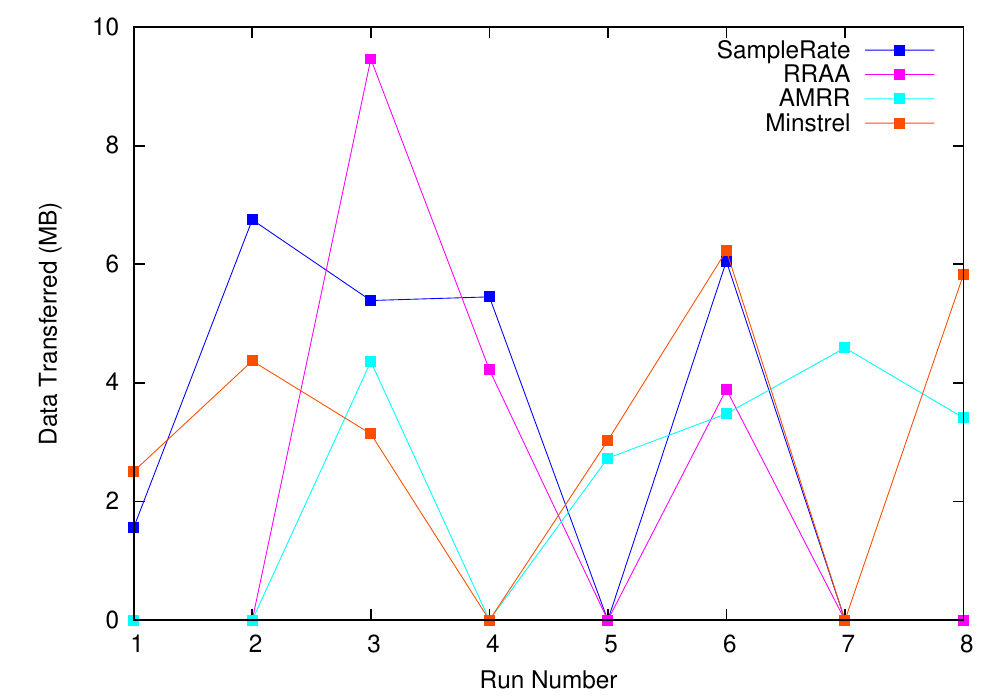 , height=2in, width=3.3in}
\caption{The variation in the amount of data transferred for RRAA, SampleRate, AMRR and Minstrel with tcp traffic }
\label{fig:var_dt_tcp}
\end{figure}

In this section, we discuss and analyze the results obtained from our experiments. Firstly, we observed significant variations in the amount of data transferred across runs for a fixed set of parameters, as shown in Figures \ref{fig:var_dt} and \ref{fig:var_dt_tcp}. For TCP traffic, with each of the rate adaptation algorithms, there were some runs in which no data was transferred. Even for UDP traffic, there were appreciable differences in the amount of data transferred. For CBR traffic of 30Mbps the least data transferred in SampleRate was 19.7\% of the most data transferred in a run. And in the case of RRAA, the least data transferred was 35.5\% of the most data transferred in a run. For AMRR, the least data transferred was 33.6\% of the most data transferred in a run. Whereas in a run of Minstrel, no data was transferred. In the section below, we discuss the connection setup protocols which were the major source of these variations.\newline

\subsection{Connection Setup Protocols}
Table \ref{tab:bitrate} shows the sequence of control messages(in descending order) that need to be sent before the transmitter can start sending useful data. This is a lengthy procedure as can be noticed from Table \ref{tab:bitrate}. Recovering from the loss of an individual control message is handled by a variety of different mechanisms across different networking layers. In a lossy environment, like the entry phase, all of these mechanisms must function well together for quick connection setup; something not easily achieved using existing protocols \cite{hadler}. Lengthy AP selection, ARP timeouts, MAC management timeouts, application initialization delay and TCP connection timeouts, all cause delay and variations in the amount of data transferred. Hadaller et al. \cite{hadler} have discussed the impact of each of these protocols on the TCP connection. \newline
In our experiments, the protocols that had a profound impact on the data connection included the lengthy AP selection procedure, ARP timeouts and TCP connection timeouts. The causes of lengthy AP selection have been discussed extensively \cite{hadler}. However the sources of ARP timeouts and TCP connection timeouts are in some way different from that caused by other control messages. Unlike other control messages, which are sent at 1Mbps, the MAC bit rate at which the ARP packets and TCP SYN, SYN+ACK  and ACK packets are sent is determined by the underlying rate adaptation algorithm. Table \ref{tab:bitrate} characterizes the different control messages on the basis of the MAC bit rates at which they can possibly be sent. In the section below, we discuss the three major sources of delay in our experiments.
\begin{table}[t]
\centering
\caption{MAC bit rates for different types of packets}
\label{tab:bitrate}
\begin{tabular}{|l | l|l|}
\hline
\textbf{Packet Type} & \textbf{Bit rate} & \textbf{Description} \\
\hline
\hline
Beacons & 1Mbps & Broadcast  \\
& & by the AP \\ \hline
Probe  & 1Mbps & Broadcast  \\
Request & & by the AP \\ \hline
Probe  & 1Mbps & Sent by the AP \\
Response& & to the Client \\ \hline
Association  & 1Mbps & Sent by the Client\\
Request & &  to the AP \\ \hline
Association  &	1Mbps &	Sent by the AP\\
Response & & to the Client \\ \hline
Authentication  & 1Mbps & Sent by the Client\\
Request & & to the AP \\ \hline
Authentication  &	1Mbps &	Sent by the AP\\
Response & & to the Client \\ \hline
ARP request & 1Mbps &	Sent by the Sender \\
& & to the Receiver \\ \hline
ARP response &	1-54 Mbps & Selected by the \\
& & receiver side \\
& & rate adaptation algorithm \\ \hline
TCP SYN & 1-54 Mbps & Selected by the \\
& & sender side \\
& & rate adaptation algorithm \\ \hline
TCP SYN+ &	1-54 Mbps & Selected by the \\
ACK& &  receiver side \\
& & rate adaptation algorithm \\ \hline
TCP ACK &	1-54 Mbps & Selected by the\\
& & sender side \\
& & rate adaptation algorithm \\ \hline
\end{tabular}
\end{table}

\subsubsection{AP Selection}
Before an 802.11 MAC connection is attempted, a client must decide which access point to connect to. A client first locates all available access points by performing one or both of (1) passive scan, which involves scanning all channels and listening for beacon messages, (2) active scan, which involves the client sending the probe requests and waiting for probe responses from all available access points. Our cards performed both simultaneously. The scanning process continues until the client locates an access point it wishes to connect to. In our case, the client has pre-configured AP ESSID. However still there was considerable amount of delay in the client getting associated with the AP, because of the AP scanning procedure. In our UDP runs this was the primary source of variations in the amount of data transferred, because we did not have to deal with TCP connection setup.

\subsubsection {ARP Timeouts}
The access point must perform an ARP lookup of the client's MAC address based on the destination IP address requested by the application. Lost ARP messages are retransmitted after one second. However, unlike \cite{hadler}, in our experiments the ARP timeouts were a significant source of delay. This is because ARP responses, the MAC bit rate of which are selected by the receiver side rate adaptation algorithms, are initially sent at high rate of 54Mbps. These rates usually fail causing several successive ARP timeouts, each lasting 1s. Given that the connection time is already small this severely reduces the time available for data transmission. Consider the case of Run 1 of RRAA with TCP. In this particular run, there were  4 ARP timeouts, each lasting  1s, because of the failure of 4 ARP responses. The reason was that these responses were sent at MAC bit rates of 54,48,36 and 24Mbps respectively in the early part of the connection, when the channel conditions were not good enough to support these bit rates. Furthermore, in our UDP runs we had fixed the receiver to a bit rate of 1M, in order to avoid ARP timeouts.

\subsubsection {TCP Connection Timeouts}
Similar to ARP responses, the MAC bit rates of the TCP SYN, SYN+ACK, ACK are selected by the sender and receiver's rate adaptation algorithms. And all algorithms use a high rate of 54Mbps to send these packets, as a result of which they have a high probability of getting lost. Losing TCP control messages is even more costly since TCP SYN timeout is about 3s. As a result there were several runs in which no data or very little data was transferred as shown in Figure \ref{fig:var_dt_tcp}. Consider the case of Run 7 of SampleRate. In this particular run, there were 5 TCP timeouts, each lasting 3s. Some of these timeouts occurred because the TCP SYN packet was sent at a bit rate which was too high. One of these timeouts occurred because the TCP ACK was sent at bit rate which was too high for the current channel conditions.

\subsubsection {Overestimation of Initial MAC Bit Rates}
All four algorithms that we tested over estimated the initial MAC bit rates. All of them used an initial MAC bit rate of 54Mbps. As a consequence, there were frequent ARP timeouts, TCP connection timeouts and the initial data packets were also lost.

\subsection {Fixed Rates}
In this section, we discuss the results from the experiments in which the MAC bit rate was fixed during the entire duration of a run. We performed such experiments for all of the twelve MAC bit rates available in 802.11g. The motivation behind performing such experiments was to assess the performance of each of the fixed rate, which in turn would allow us to better evaluate the performance of each of the rate adaptation algorithms.

\subsubsection{DSSS vs OFDM rates}
\begin{figure}[t]
\centering
\epsfig{file=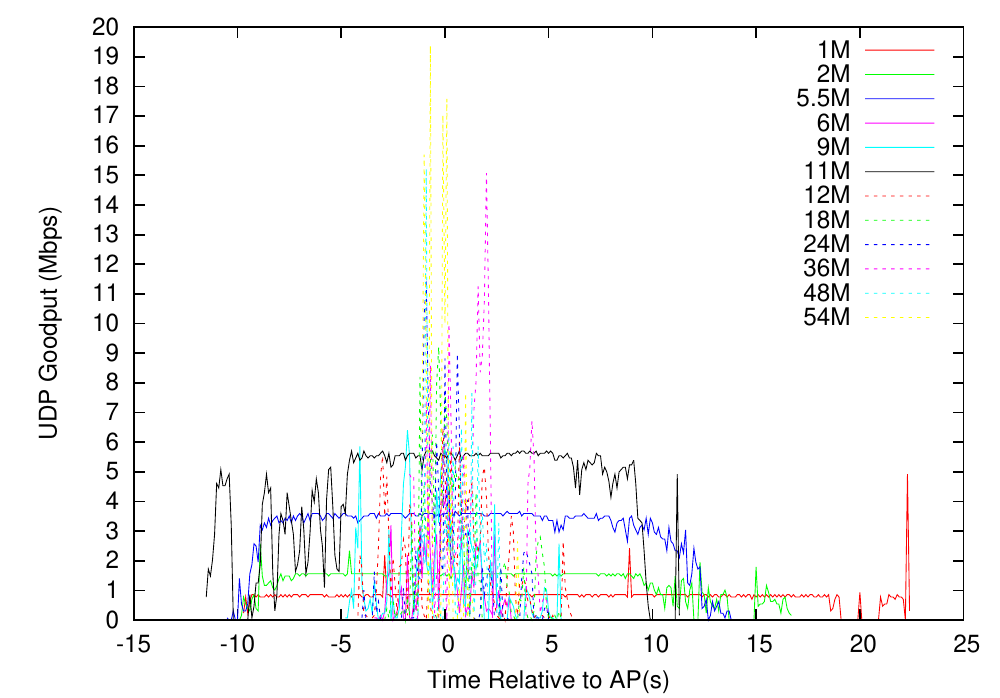 , height=2in, width=3.3in}
\caption{Supremum goodput for each fixed rate across 5 runs}
\label{fig:sup_goodput}
\end{figure}

\begin{figure}[t]
\centering
\epsfig{file=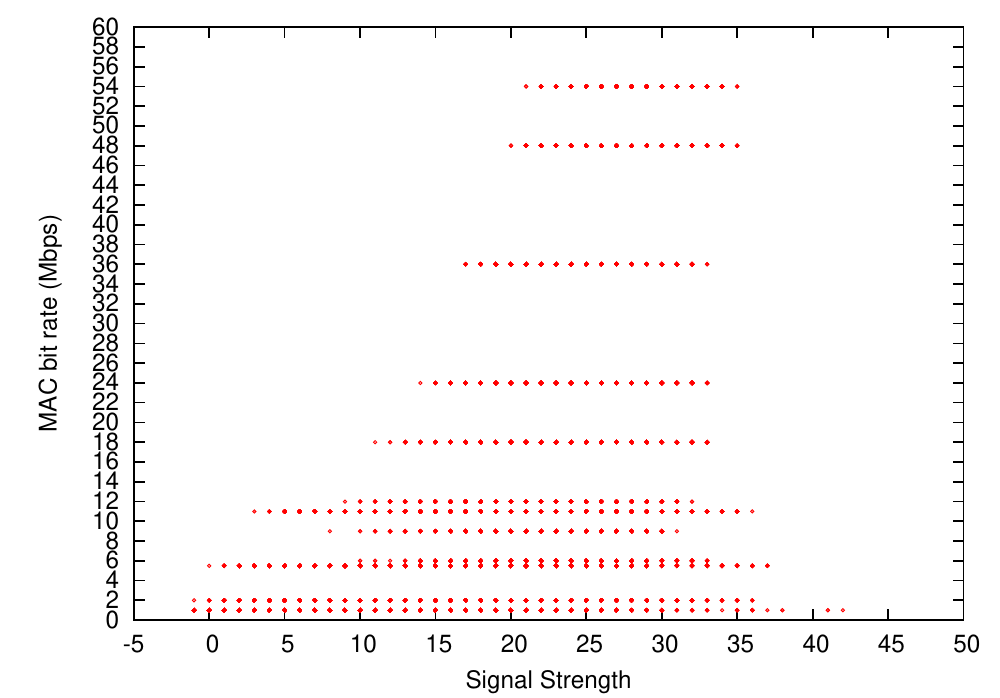 , height=2in, width=3.3in}
\caption{Receive signal strength values corresponding to each MAC bit rate}
\label{fig:rssi}
\end{figure}

Figure \ref{fig:sup_goodput} shows the supremum goodput plots for each fixed rate over 5 runs (with 30 Mbps CBR traffic over UDP). 1,2,5.5 and 11Mbps which use DSSS modulation scheme tend to have a longer transmission range as compared to 6,9,12,18,24,36,48 and 54Mbps which use OFDM modulation scheme. Secondly, in each of these two set of rates, the higher the rate, the smaller the transmission range. In fact, 54Mbps in our settings had a transmission range of approximately 50m. As shown by Figure \ref{fig:sup_goodput}, the rate that performed the best was 11Mbps, followed by 5.5Mbps. The reason being that these rates were able to maintain a relatively high steady throughput performance over a long period of time. Rates like 54,48 and 36Mbps did provide very high throughput but for an extremely short period of time. \emph{This shows that rates like 11Mbps and 5.5Mbps which have a longer transmission range and provide a relatively high steady throughput, are the ones which are going to succeed most frequently in a lossy environment like the one that we have considered. Whereas higher rates like 54,48 and 36Mbps should be used very carefully because of their low probability of success.  \newline}
Figure \ref{fig:rssi} shows the range of RSSI values, needed to decode frames at different MAC bit rates, based on our fixed rate experiments. An interesting insight that can be gained from this graph is that 11Mbps has a lower RSSI threshold than 6Mbps and 9Mbps. This is one of the reasons why 11Mbps had a higher transmission range than 6Mbps and 9Mbps. \emph{Consequently 6Mbps and 9Mbps should never be selected by rate adaptation algorithms if 11Mbps is available. Another inference that can be drawn is that control messages like ARP packets, TCP SYN, TCP SYN+ACK and TCP ACK packets should never be sent at OFDM rates, which have smaller transmission range and require higher RSSI for demodulation, since losing these packets can have an adverse affect on the overall connection.}\newline\newline

\subsubsection{Link Layer Acknowledgement Rates}
Another interesting aspect is that the 802.11 standard specifies that the an ACK frame should be transmitted at the highest rate in the basic rate set, that is less than or equal to the transmission rate of the data frame it is acknowledging. Lets call such ACK transmission rate the default ACK rate.
For example, the 802.11g basic rate set is 1, 2, 5.5, 11, 6, 12, 24 Mbps. So if a data frame
is transmitted at 18 Mbps, the default rate of the corresponding ACK frame is 12 Mbps
In practice, Madwifi allows two different transmission rates for ACK frames as listed in the Table \ref{tab:ack_rates} \cite{prateek}.

\begin{table}[t]
\centering
\caption{ACK rates corresponding to different data rates}
\label{tab:ack_rates}
\begin{tabular}{|c|c|c|}
  \hline
  \textbf{Data Rate} &  \textbf{Low ACK Rate} & \textbf{High ACK Rate} \\ \hline \hline
  1 & 1 & 1 \\ \hline
  2 & 2 & 2 \\ \hline
  5.5 & 2 & 5.5 \\ \hline
  6 & 6 & 6 \\ \hline
  9 & 9 & 9 \\ \hline
  11 & 2 & 11 \\ \hline
  12 & 6 & 12 \\ \hline
  18 & 6 & 12 \\ \hline
  24 & 6 & 24 \\ \hline
  36 & 6 & 24 \\ \hline
  48 & 6 & 24 \\ \hline
  54 & 6 & 24 \\
  \hline
\end{tabular}
\end{table}

\begin{figure}[t]
\centering
\epsfig{file=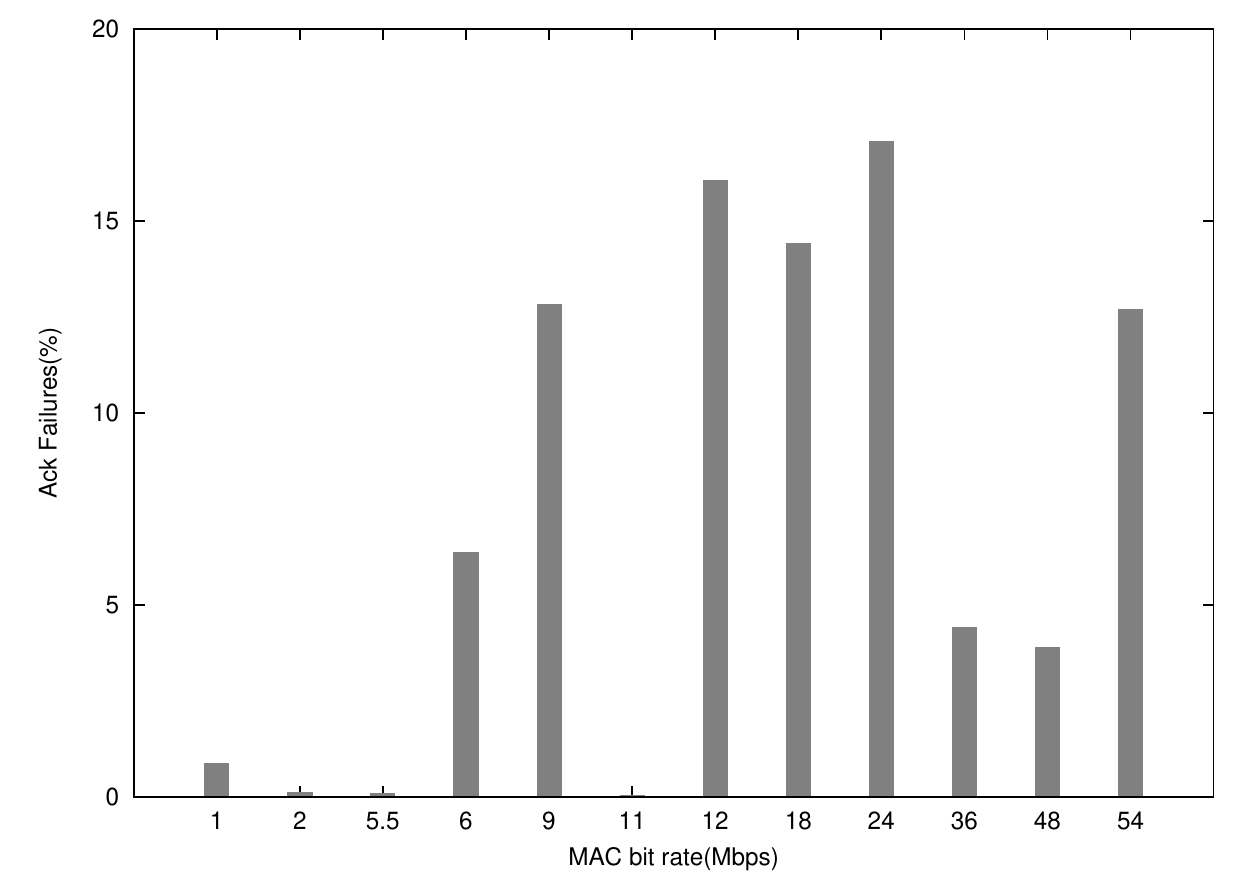 , height=2in, width=3.3in}
\caption{Acknowledgement failures for each fixed rate as a percentage of total frames received successfully(averaged over five runs)}
\label{fig:ack_fail}
\end{figure}

In our experiments, we observed that for DSSS rates, low ACK rates were selected whereas for OFDM rates, high ACK rates were selected. When an ACK gets lost, the sender side MAC assumes that the frame was lost, so it invokes an exponential backoff before retransmitting the frame. In a our enviornment, high ACK rates tended to decrease the overall probability of success of a frame. Figure \ref{fig:ack_fail} shows the ACK failures corresponding to each MAC bit rate as a percentage of total frames successfully received. 17.1\% of frames sent at 24Mbps, that were received successfully by the receiver, were assumed to be lost by the sender because their ACKs failed. This is because the ACKs for 24Mbps were sent at high rate of 24Mbps. The probability of success of a data frame sent at high rate of 24Mbps is relatively low because of its smaller transmission range and the fluctuations in RSSI. As a consequence, when the ACK is also sent at 24Mbps, the overall probability of success of the frame is reduced even further. Similarly OFDM rates had a relatively high percentage of ACK failures. On the other hand, DSSS rates had a very low percentage of ACK losses because their ACKs were sent at low rates of 1Mbps and 2Mbps, which have a longer transmission range and supports lower RSSI for demodulation. \emph{This again shows that OFDM rates should be used very carefully because their ACKs are also sent at relatively high rates, and hence their overall probability of success at any stage is smaller or equal to that of DSSS rates.}

\subsubsection{Supremum Goodput and MAC Bit Rate across all Fixed Rates}
\begin{figure}[t]
\centering
\epsfig{file=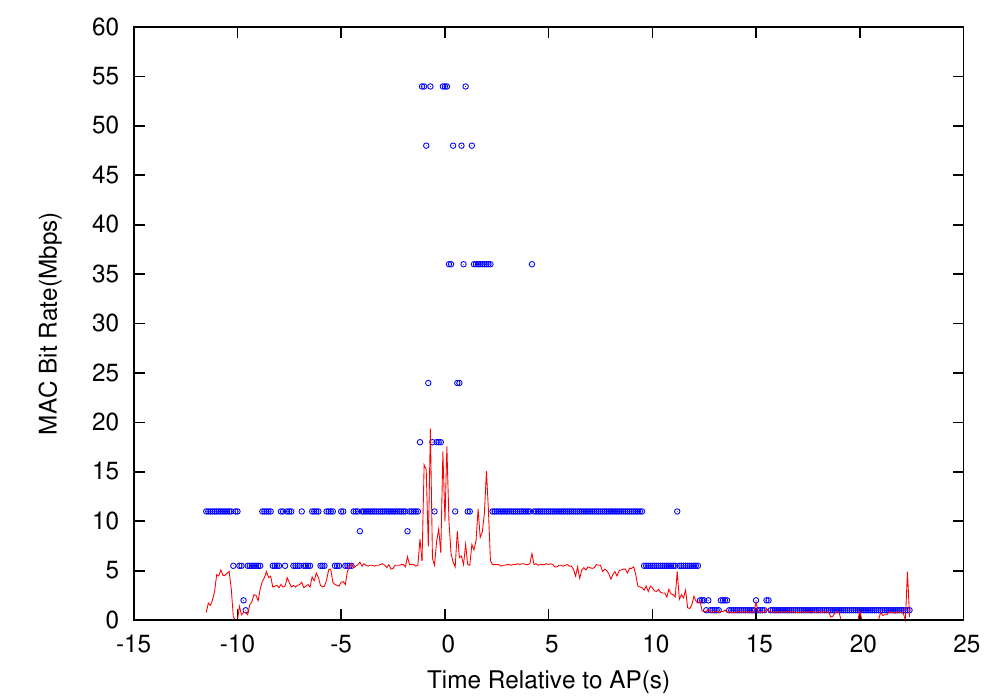 , height=2in, width=3.3in}
\caption{The supremum goodput across all fixed rates, and the corresponding MAC bit rate}
\label{fig:sup_goodput_all}
\end{figure}

Figure \ref{fig:sup_goodput_all} shows supremum goodput of all fixed rates and the corresponding MAC bit rates. This supremum goodput was found by calculating the maximum goodput in each 0.1s interval over all fixed rates. The MAC bit rates, shown, give us a good idea of a close to ideal MAC bit rate strategy. Under such a strategy the total data transferred is 16.6 MBytes. This is 111\% of the maximum data transferred by any rate adaptation algorithm in any run.

\subsection{Rate Adaptation Algorithms}
The fixed rate analysis in the earlier section has provided us with useful insights on rate selection in the vehicular setting. As we move to rate adaptation algorithms, we briefly summarize some of the key insights gathered from fixed rate analysis. \newline
OFDM and DSSS rate have different characteristics in terms of their transmission range and RSSI threshold for demodulation. DSSS rates like 11Mbps and 5.5Mbps provide steady throughput performance whereas rates like 54,48,36,24,18 and 12Mbps do provide high throughput but only for a short time interval. Generally the channel conditions(as shown by the RSSI value) are not good enough to support these OFDM rates. Secondly 6,9Mbps have lower transmission range and RSSI threshold for demodulation than 11Mbps, and hence should never be selected.
\newline
\begin{figure}[t]
\centering
\epsfig{file=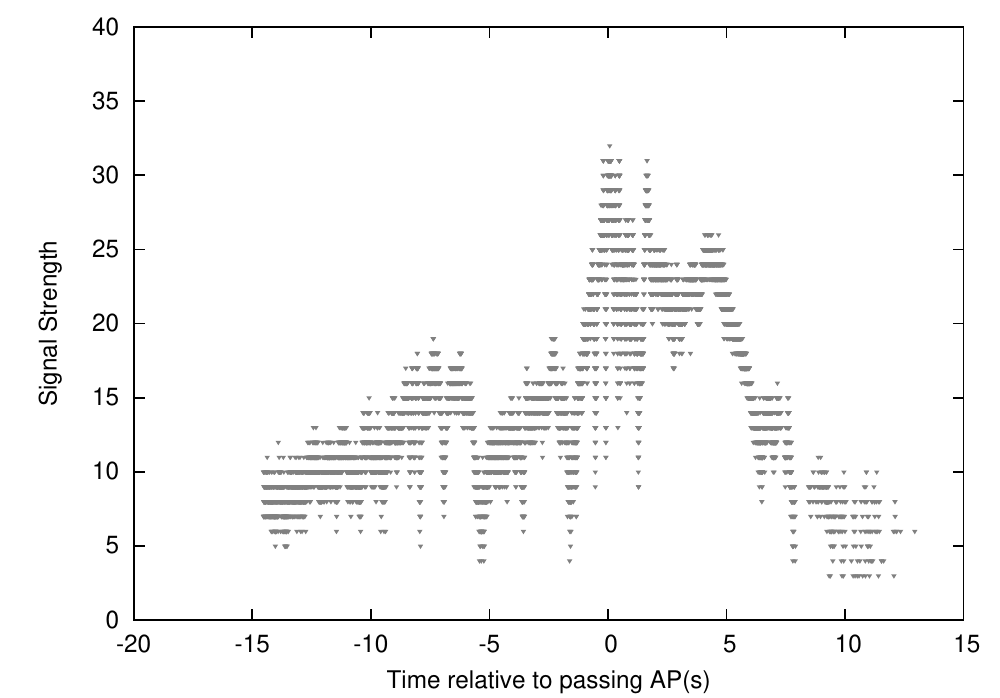 , height=2in, width=3.3in}
\caption{A typical Signal Strength Profile in our settings}
\end{figure}
We tested RRAA, SampleRate, AMRR and Minstrel with CBR traffic(over UDP) of 10Mbps and 30Mbps as well as with TCP bulk traffic. Each of these algorithms have an estimation window; an interval of packets or time which is used to predict the next bit rate. In the vehicular context, the channel conditions change quite rapidly because of the mobility of the vehicle. For instance, in our case, in 10s the car traveled approximately 150m, a distance through which channel conditions change tremendously as shown by Figure 10. As a consequence, if the estimation window is too large, it might not represent the channel conditions accurately.\newline
All of rate adaptation algorithms either employ sequential rate switching or best transmission rate policy. In sequential rate switching the problem is that if a rate is selected which is far away from a suitable rate, it is going to take the algorithm some time to converge to the suitable rate. And in that time channel conditions might have changed significantly, rendering the selected rate inappropriate.\newline
Some of the above algorithms like SampleRate, AMRR and Minstrel also use probe packets. These packets are used to send packets at rates other than the current rate to assess the performance of other bit rates. However in a lossy environment like the one that we have considered, the likelihood of a packet getting lost is reasonably high. So if a probe packet at a bit rate gets lost, it is likely that rate might not be selected for some time even if it was one of the most suitable rates.\newline
Every algorithms starts of by selecting some initial rate. In the case of the above algorithms, they start off by selecting a high initial bit rate irrespective of the channel conditions as shown in Table 3. High rates as discussed in the earlier sections are viable only when the channel conditions are extremely good, however often at the start of the connection the channel conditions are not good enough to support such high bit rates.\newline
In the section below, we discuss and analyze the results from these experiments.

\subsubsection{Rate Adaptation Algorithms with UDP}
\medskip
\textbf{RRAA}\newline
This algorithm uses short-term loss estimation of 802.11 frames (in a window of tens of frames)
to opportunistically guide rate adaptation. RRAA has two modules -- RRAA-BASIC and an adaptive RTS filter. The RRAA-BASIC contains the lost estimation and the rate change algorithm, whereas the adaptive RTS filter selectively turns on RTS/CTS exchange to suppress collision losses. In the scenario that we consider there is no interference from other 802.11 sources, therefore we only use the RRAA-BASIC module.  The code for the implementation for RRAA has been taken from \cite{ramachandran}. In this implementation, RRAA is invoked every 200ms or after 40 packets have been received (defined as an interval), and the algorithms uses loss rate estimated in the last interval for rate adaptation.\newline
As discussed in the previous section, RRAA uses an initial MAC bit rate of 54Mbps. As mentioned above, it is only every 200ms or after 40 packets that a rate change decision will take place. Hence, if a rate is too high with respect to the channel conditions, it will take a considerable amount of time for the appropriate to be selected, causing the UDP goodput to go down initially. Figures 11, 12 and 13 shows a few such runs which illustrate this point. Secondly, RRAA did try to send frames at 6, 9Mbps when 11Mbps would have provided a better rate choice(see Figures 11 and 12).\newline
\begin{figure}[t]
\centering
\epsfig{file=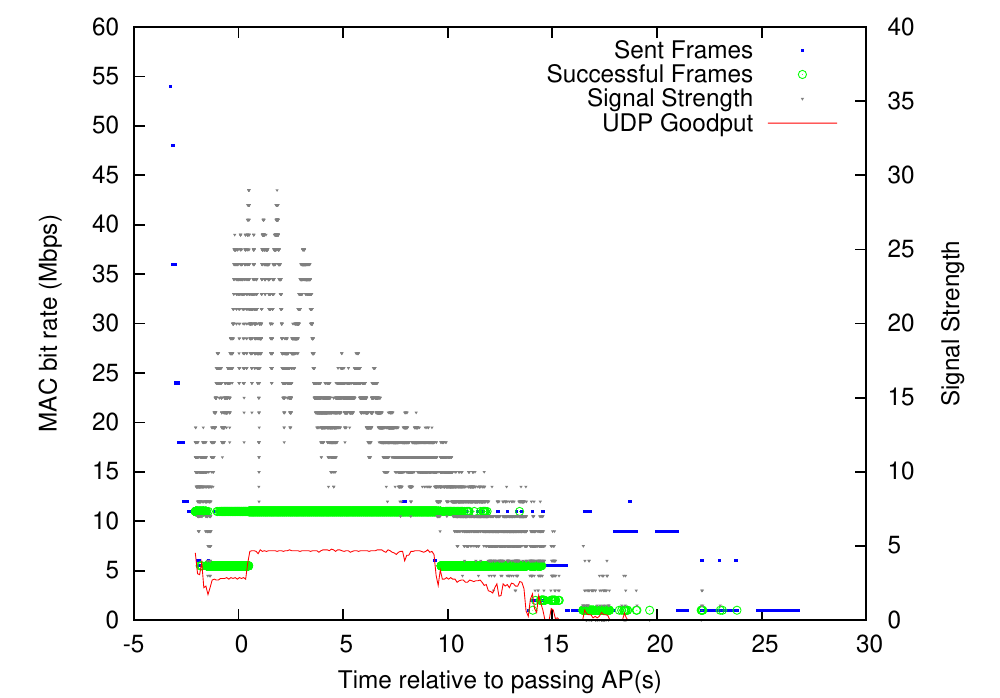 , height=2in, width=3.3in}
\caption{UDP goodput, sent frames and successful frames against time relative to passing AP for a RRAA run with CBR traffic 10Mbps}
\end{figure}
\begin{figure}[t]
\centering
\epsfig{file=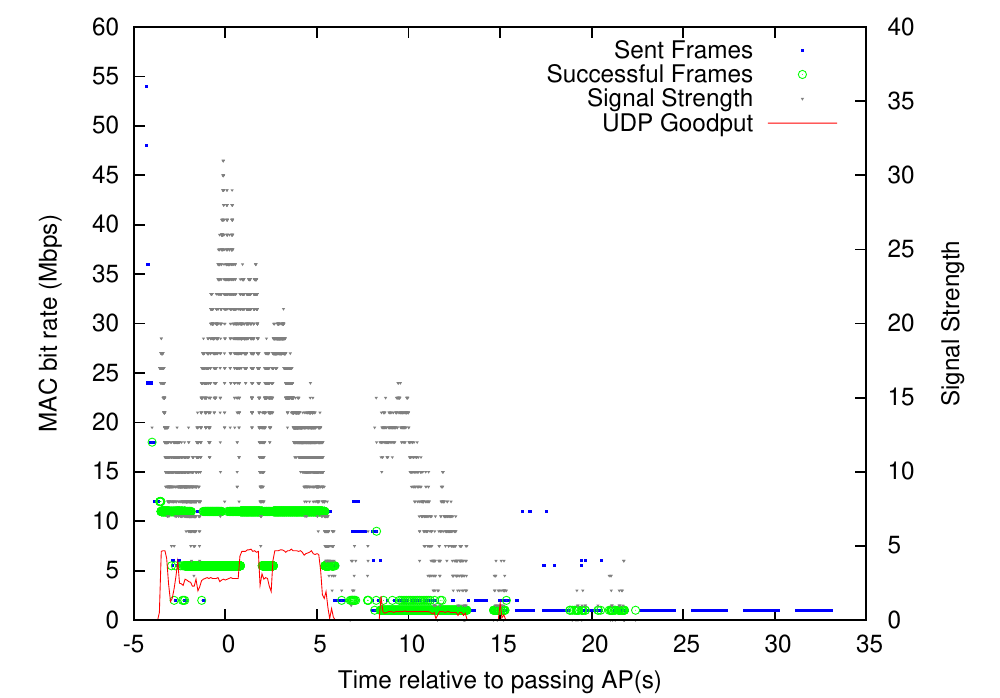 , height=2in, width=3.3in}
\caption{UDP goodput, sent frames and successful frames against time relative to passing AP for a RRAA run with CBR traffic 30Mbps}
\end{figure}
Thirdly, the algorithm does not take into account the characteristics of different rates i.e. the DSSS rates behave differently from OFDM rates, hence the two sets of rates should be treated differently. RRAA did try to use OFDM rates even when RSSI indicated that these rates would fail, resulting in UDP goodput drop as shown in (see Figure 11 and 12).\newline
The runs of RRAA in which it was able to stabilize the rates of 11Mbps and 5.5Mbps were the ones in which most data was transferred. Figure 13 shows such runs of RRAA in which considerable amount of data was transferred. If we compare Figure 13 with Figure 6, containing the supremum of the supremum of all fixed rates, we realize that the key difference is that the ideal rate strategy indicates that close to the AP, when channel conditions, as indicated by the SNR are good, high rates like 54, 48 and 36Mbps can be used. However RRAA wasn't able to adapt to channel conditions and send frames at these high rates.\newline
\begin{figure}[t]
\centering
\epsfig{file=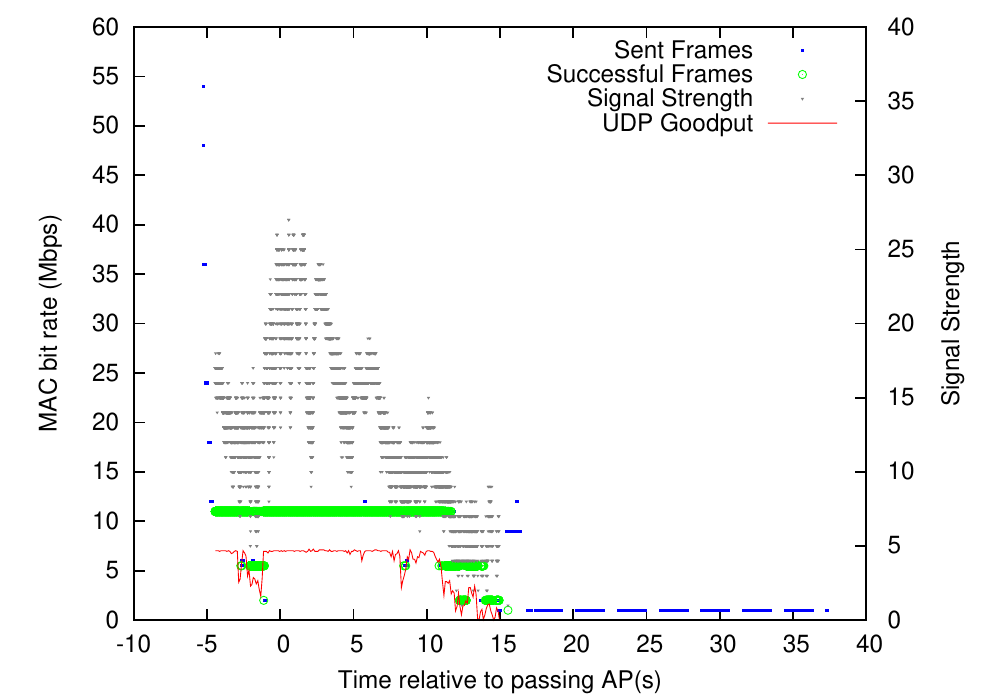 , height=2in, width=3.3in}
\caption{UDP goodput, sent frames and successful frames against time relative to passing AP for a RRAA run with CBR traffic 30Mbps}
\end{figure}
\newline
\textbf{SampleRate}\newline
SampleRate transmits packets and periodically (every 10th packet) picks up a random rate other than the current one and collects the statistics. Average transmission time plays a major role in the working
of this algorithm. Sample stops probing the bit-rates that have a poor history, it stops
sampling the bit-rates out of the list available with 4 successive failed transmissions.
The average transmission time is calculated using packet size, the bit-rate and the
number of retries needed to transmit the packet. Sample chooses to transmit data at the
rate which it predicts to have the lowest average transmission time including the time
needed for any retransmissions that are needed.\newline
SampleRate uses an estimation window of 10s i.e. the statistics of only those packet are considered which were sent in this time window. For vehicular settings this estimation window is too large, since channel conditions change extremely rapidly. In 10s, our vehicle travels approximately a distance of 150m, and in this distance channel quality changes tremendously. A large estimation window affects the ability of SampleRate to react to changing channel conditions. Therefore it was very rare for SampleRate to jump to higher rates from lower rates even when the channel conditions were good enough to support high rates (this can be seen in Figures 14 and 15). Once the rate is dropped initially to the 1Mbps, it is very difficult to immediately switch to the higher more effective rates (5.5 or 11Mbps). The switch can take place only if the probe packet at 5.5 pr 11 succeeds. This probing however is too infrequent (1 packet in 10). If the probe packet fails, the rate is not selected until this rate is again selected for the probe packet later. Moreover in a lossy environment like the one that we have considered, the likelihood of packet getting lost is relatively high, hence it is possible that few rates become out of favor for some time.
\newline
\begin{figure}[t]
\centering
\epsfig{file=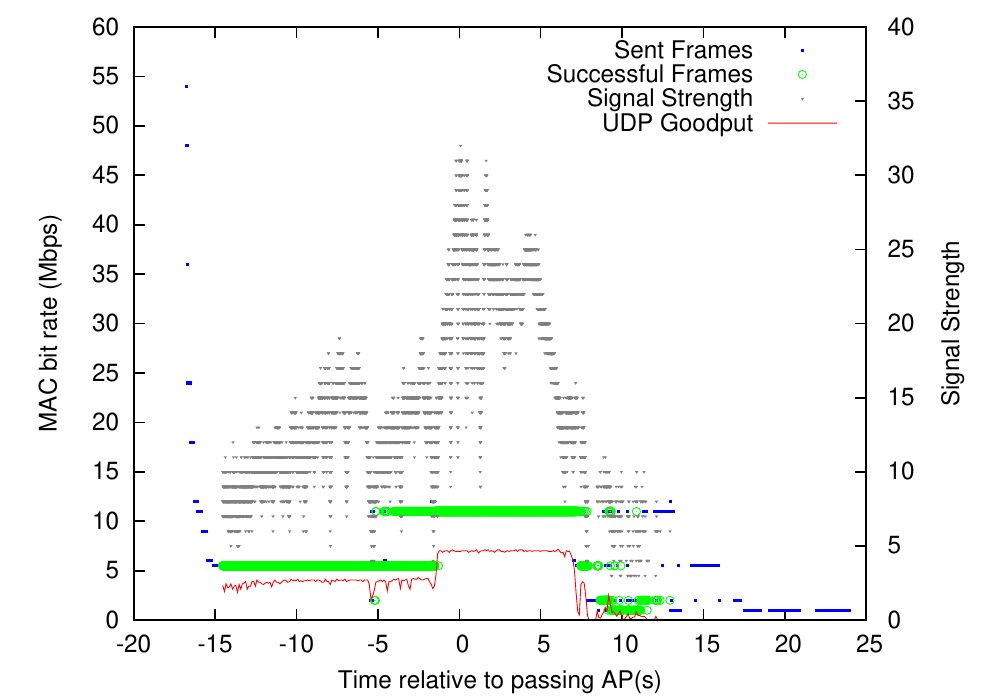 , height=2in, width=3.3in}
\caption{UDP goodput, sent frames and successful frames against time relative to passing AP for a SampleRate run with CBR traffic 10Mbps}
\end{figure}

\begin{figure}[t]
\centering
\epsfig{file=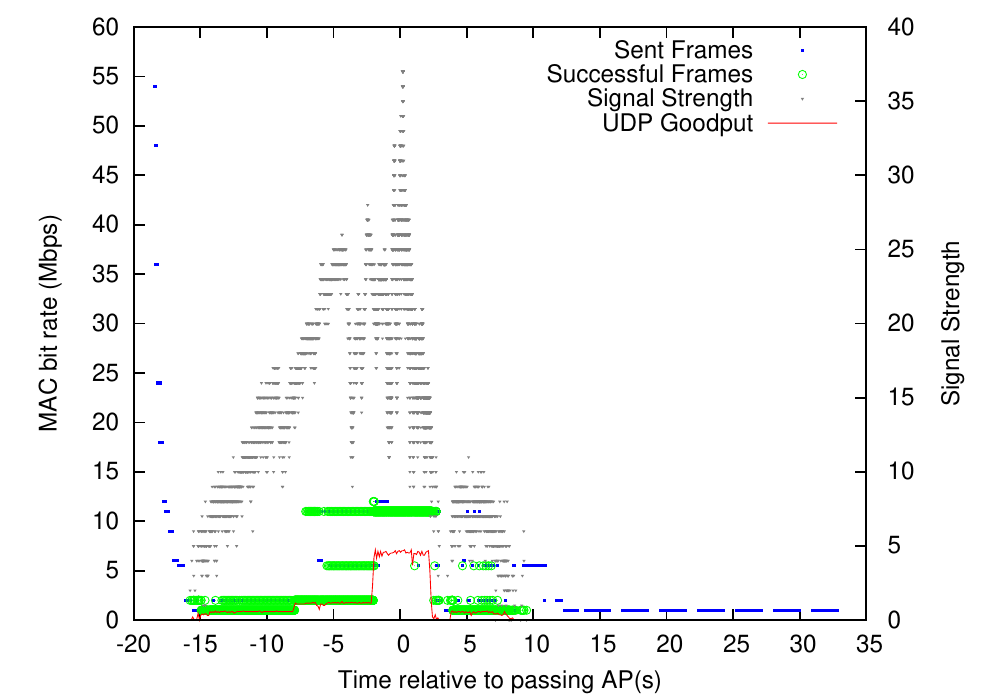 , height=2in, width=3.3in}
\caption{UDP goodput, sent frames and successful frames against time relative to passing AP for a SampleRate run with CBR traffic 30Mbps}
\end{figure}

\noindent \textbf{AMRR}\newline
AMRR tries to capture the short-term variations by selecting rate whose PER is low enough such that the number of retransmissions are low. AMRR has an estimation window of 1 second. It keeps track of the retries corresponding to the primary rate and if these retries are less than 10\% of the distinct packets transmitted, this rate is classified as successful; otherwise if the retries are greater than 33\% of the distinct packet transmitted, the rate is marked as failed and rate is decreased. In case of success, if a certain threshold is reached, a probe packet at a higher rate is sent. If this packet fails, the rate is immediately dropped to the lower rate.\newline
AMRR uses the percentage of retries to the total packets transmitted to infer the performance of a rate algorithms. As explained above it uses the thresholds of 10\% and 33\% to infer rate increase and rate decrease respectively. However the problem is that different rates may require different thresholds. 
\newline
Another important point to note is that in a lossy environment that we have considered, most rate suffer decent number of losses. This means if the threshold are fixed to some low value, it might prevent some of the high rate from being selected.
\newline
AMRR uses probe packets to assess bit rates other than the current one. However in a lossy environment like the one that we have considered, it makes rate switching on the basis of probe packet difficult. This because it is likely that the probe packet will fail, causing the rate to become out of favor. As a consequence we saw that in the case of AMRR, there weren't frequent rate changes.
\begin{figure}[t]
\centering
\epsfig{file=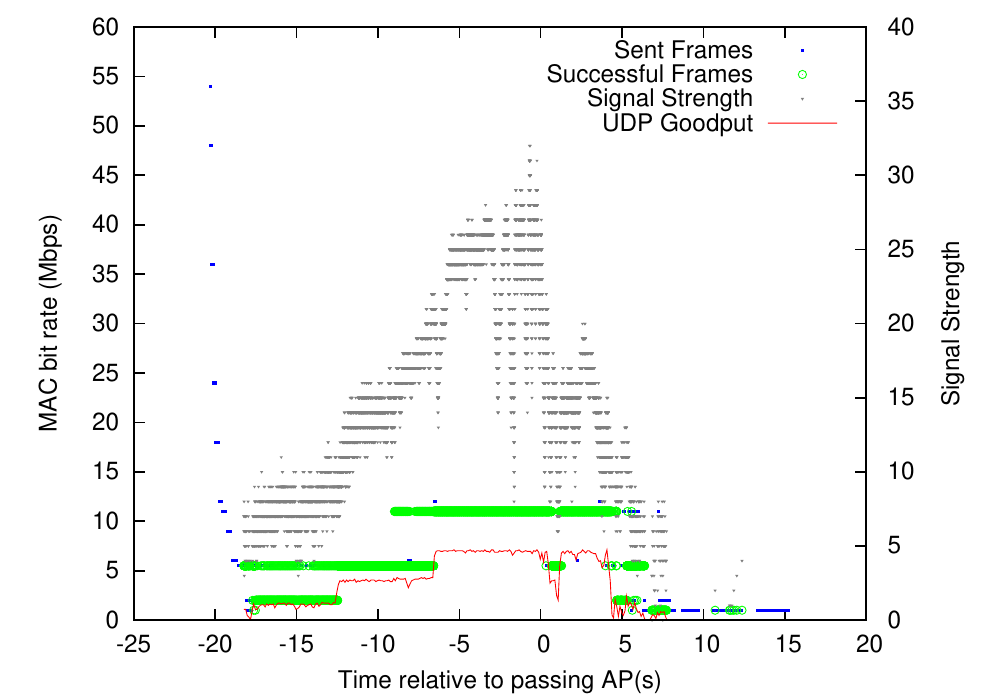 , height=2in, width=3.3in}
\caption{UDP goodput, sent frames and successful frames against time relative to passing AP for a AMRR run with CBR traffic 30Mbps}
\end{figure}
\begin{figure}[t]
\centering
\epsfig{file=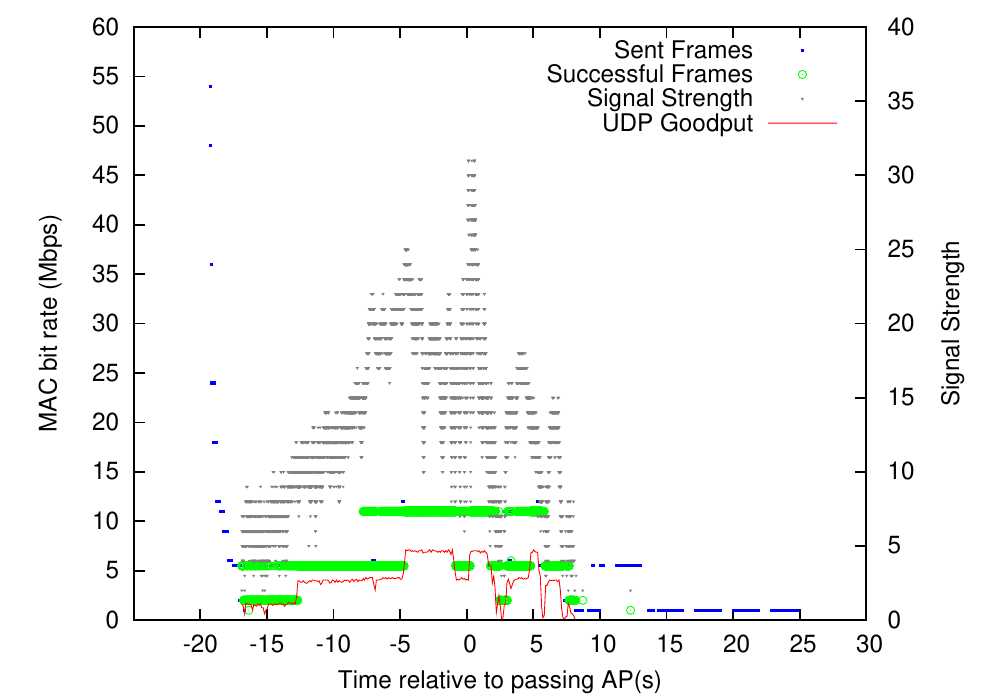 , height=2in, width=3.3in}
\caption{UDP goodput, sent frames and successful frames against time relative to passing AP for a AMRR run with CBR traffic 30Mbps}
\end{figure}
\begin{figure}[t]
\centering
\epsfig{file=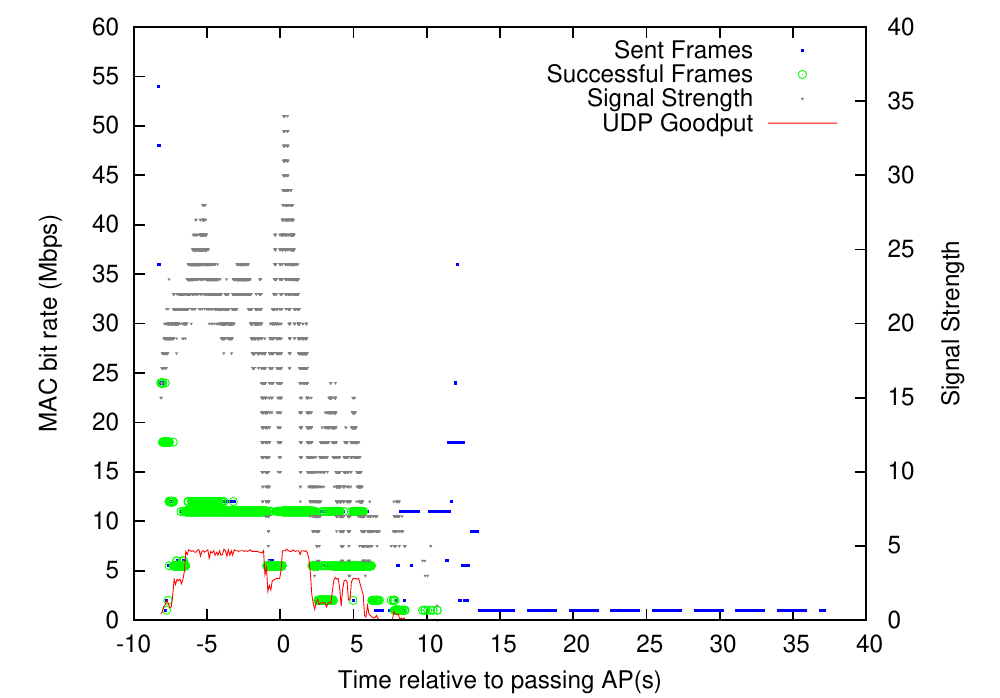 , height=2in, width=3.3in}
\caption{UDP goodput, sent frames and successful frames against time relative to passing AP for a AMRR run with CBR traffic 10Mbps}
\end{figure}
\newline
\\
\textbf{Minstrel}\newline
The basic idea behind this algorithm is transmit at different rates whenever possible other than the current one and switch to the rate that provides the best opportunity for maximum throughput.
\newline
Minstrel is a EWMA based algorithm. It uses similar ideas as used by the Sample algorithm. It uses a formula to compute the successfulness of packet transmission. This measure of successfulness is used to adjust the
transmission speed to the optimum level. It dedicates a particular percentage of data
packets to be transmitted at different rates other than the current one and is set to 10\% in the
default configuration and the algorithm fires at a definite time interval which is set at
100 milliseconds (10 times per second) in the default configuration. Minstrel keeps track of statistics for 100ms and updates the rate in its retry chain. The probability of success is associated with every rate and the rate which can achieve the best throughput is selected. However we have disabled multi-rate retry and hence the whole retry chain cannot be executed in our case. Furthermore MINSTREL had EWMA factor set to 75\% (meaning old results are paid significant attention). This had the effect that MINSTREL tried to retain the initial best rates selected and seldom tried higher rates, exclusively. This is the case in Figures 19 and 20 and where Minstrel stabilized 5.5Mbps throughout the connection period and tried to switch to 11Mbps, but remained intact with the current rate. Thus, the high SNR period was not utilized effectively and the overall data transferred was reduced. Also in the case where the rate was initially stabilized to 1Mbps, MINSTREL was not able to switch to a more suitable higher rate, exclusively, as can be seen in Figure 21; where the initial rates were stabilized to 1Mbps and 2Mbps. In cases where MINSTREL stabilizes 11Mbps from the very beginning of the connection, MINSTREL tries to retain this rate over the entire connection period. In effect this allows large amount of data to be transferred as can be seen in the Figure 22.\newline
\begin{figure}[t]
\centering
\epsfig{file=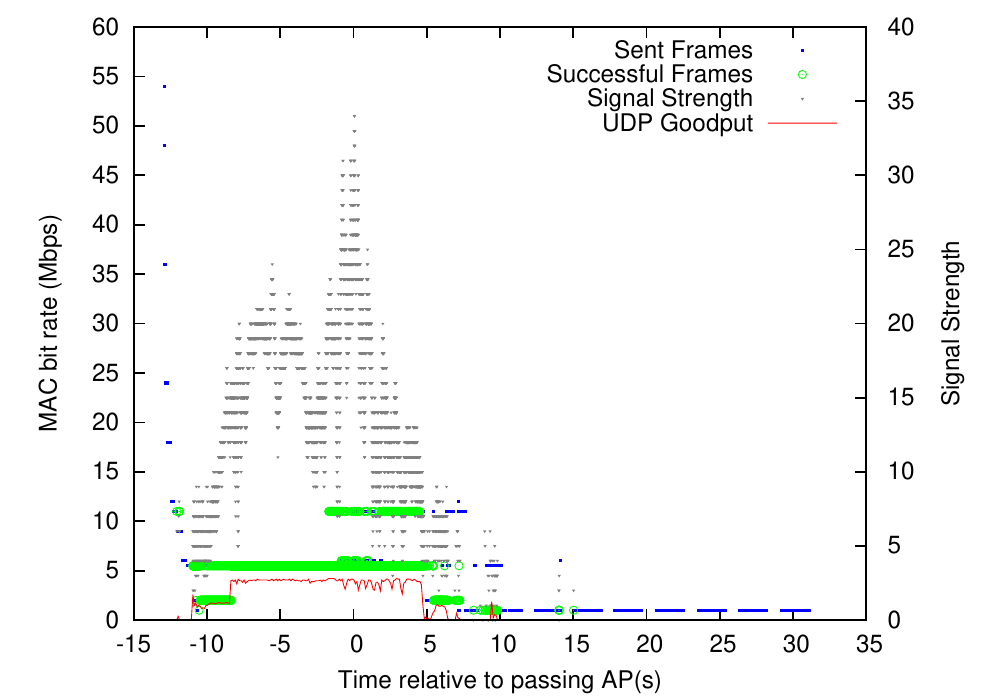 , height=2in, width=3.3in}
\caption{UDP goodput, sent frames and successful frames against time relative to passing AP for a Minstrel run with CBR traffic 10Mbps.}
\end{figure}

\begin{figure}[t]
\centering
\epsfig{file=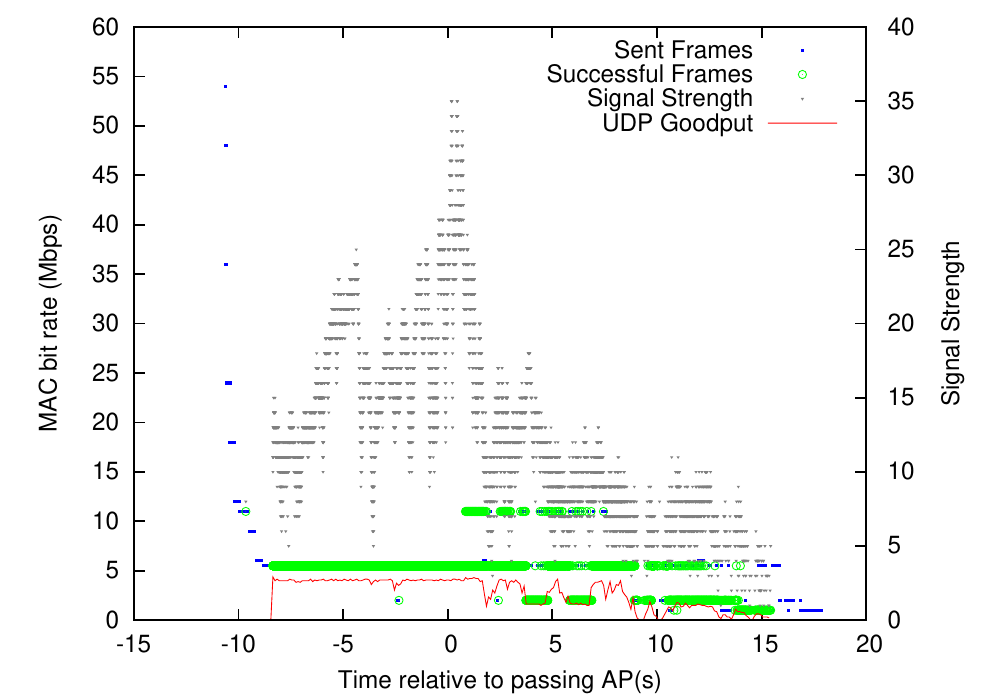 , height=2in, width=3.3in}
\caption{UDP goodput, sent frames and successful frames against time relative to passing AP for a Minstrel run with CBR traffic 10Mbps.}
\end{figure}

\begin{figure}[t]
\centering
\epsfig{file=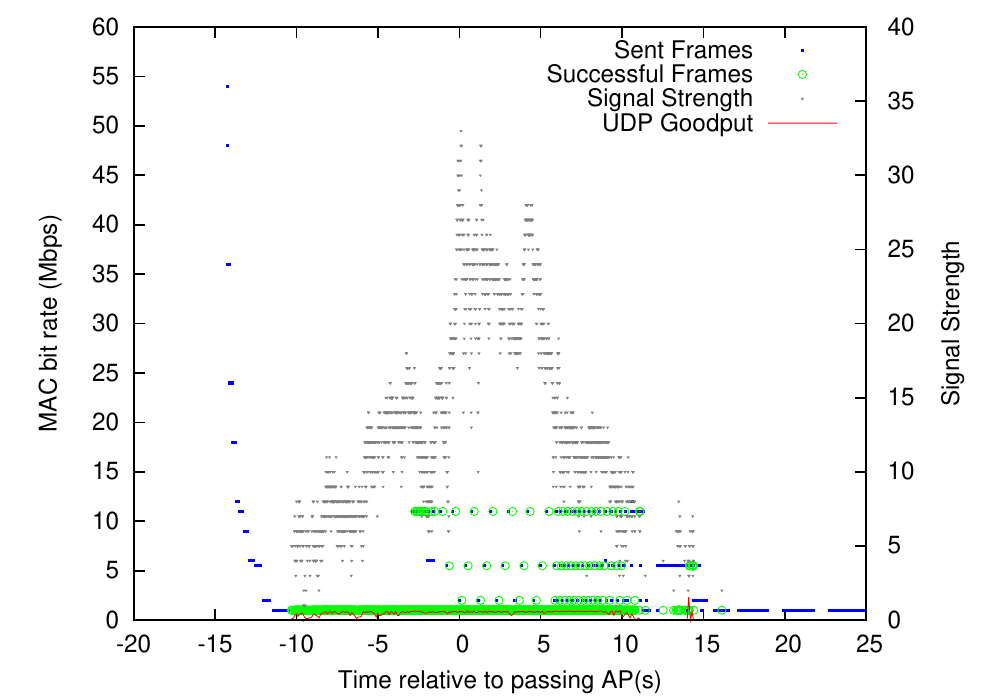 , height=2in, width=3.3in}
\caption{UDP goodput, sent frames and successful frames against time relative to passing AP for a Minstrel run with CBR traffic 30Mbps.}
\end{figure}

\begin{figure}[t]
\centering
\epsfig{file=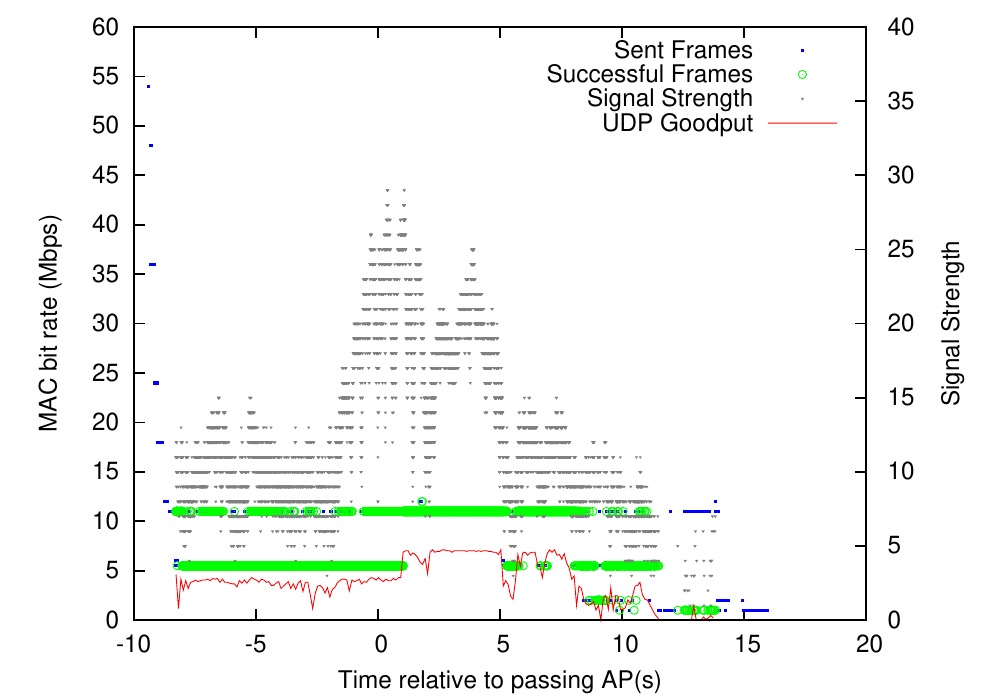 , height=2in, width=3.3in}
\caption{UDP goodput, sent frames and successful frames against time relative to passing AP for a Minstrel run with CBR traffic 30Mbps.}
\end{figure}

\subsubsection{Rate Adaptation Algorithms with TCP}
We also tested each of the four rate adaptation algorithms RRAA, SampleRate, AMRR and Minstrel with TCP traffic. In the case of TCP, the situation becomes more complex. This is because TCP employs a retransmission mechanism at the transport layer. As a consequence if a packet fails i.e. all the frame retries fail, the loss is reported to the transport layer which tries to recover from this loss by retransmitting the packet. If consecutive packet losses occur, TCP invokes exponential backoff. Typically TCP performs well with a loss rate of below 5\%, so if the TCP losses increase it is possible that TCP performance might degrade. Rate adaptation algorithms have to ensure that they select bit rates which not only try to maximize frame throughput but also ensure that TCP packet losses are minimized. \newline
In the section below we discuss the performance of each of these rate adaptation algorithms with TCP.\newline \\
\noindent \textbf{RRAA}\newline
As discussed in the previous section, RRAA uses an initial MAC bit rate of 54Mbps. As mentioned above, it is only every 200ms or after 40 packets that a rate change decision will take place. Hence, if a rate is too high, it will take a considerable amount of time for the appropriate to be selected, given that RRAA increases/decreases rates sequentially.  Consequently, some of the runs suffered from TCP connection time outs  as noted in Section 4.1.2. In fact, no appreciable TCP data could be transferred in run 1,2,5,7 and 8 of RRAA because of successive TCP timeouts. In addition, the initial TCP goodput is very low (because of losses and TCP timeouts)in some runs until an appropriate lower rate is selected (as shown in Figure 23). Losses if exposed to TCP can turn out to be extremely expensive, because TCP will then invoke exponential backoff. Figure 23 and 24 shows two such cases where because of TCP packet losses, TCP invoked exponential backoff, as a consequence very few packets were transmitted, even though the channel conditions(very close to the AP) were good enough to support several MAC bit rates. In a run shown by Figure 23, the TCP goodput drops to zero in the interval -2 to +2 seconds, this is a interval where the TCP goodput was expected to be the best. The question is why did TCP suffer from such losses? \newline
Results show that this was the result of the interplay of poor MAC bit rate selection and TCP mechanisms that resulted in such a high number of TCP losses. Lets consider the run in Figure 23, why was the rate selection poor? Well firstly RRAA selects an initial MAC bit rate of 54Mbps, which is too high with respect to the underlying channel conditions. In contrast a lower rate of 1,2,5.5 or 11Mbps might have sufficed at the start stage(as shown by our fixed rate analysis). RRAA selects an initial MAC bit rate of 54Mbps irrespective of the channel conditions. The initial rate is even more important in the context of TCP because if a TCP SYN or SYN+ACK gets lost, TCP will timeouts for 3s, as can be seen in Figure 23. Secondly the estimation window of 200ms or 40 packets is too large for vehicular settings where the channel conditions change extremely rapidly because of vehicle being in motion. As a result we see that it takes RRAA more than 1s to drop from 54Mbps to 11Mbps. Consequently if there are too many frame losses, these may well translate into TCP losses as seen in Figure 8. Thirdly as discussed in the section of UDP, RRAA does try to send frames at 6,9Mbps when 11Mbps would have provided a better(see Figure 24). Fourthly RRAA performs sequential rate switching, which means that if at any stage the selected rate is far away from the appropriate rate, it will take some time for the algorithm to converge to the appropriate rate. And in that time many frame losses might occur result in TCP losses, drop in TCP goodput. Fifth, the algorithm does not take into account the characteristics of different rates i.e. the DSSS rates behave differently from OFDM rates, hence the two sets of rates should be treated differently. As the distance from the AP increases and the channel quality decreases as indicated by the RSSI, the OFDM rates become less and less viable. Frames sent at these rates are likely to suffer many frame losses resulting in TCP losses.
\newline
The two runs of RRAA in Figures 23 and 24 show the drastic drop in TCP throughput that can be caused due to poor rate selection. An important point to note is that rate adaptation algorithms intend to maximize the frame throughput however a higher frame throughput does not necessarily lead to a higher TCP goodput. Consider that a MAC bit rate selection algorithm selects 54Mbps, which maximizes the frame throughput over all the fixed rates, however it suffers from 50\% loss rate. It is likely that losses will be exposed to TCP in this case resulting in TCP invoking exponential backoff as seen in Figure 8.\newline \\

\begin{figure}[t]
\centering
\epsfig{file=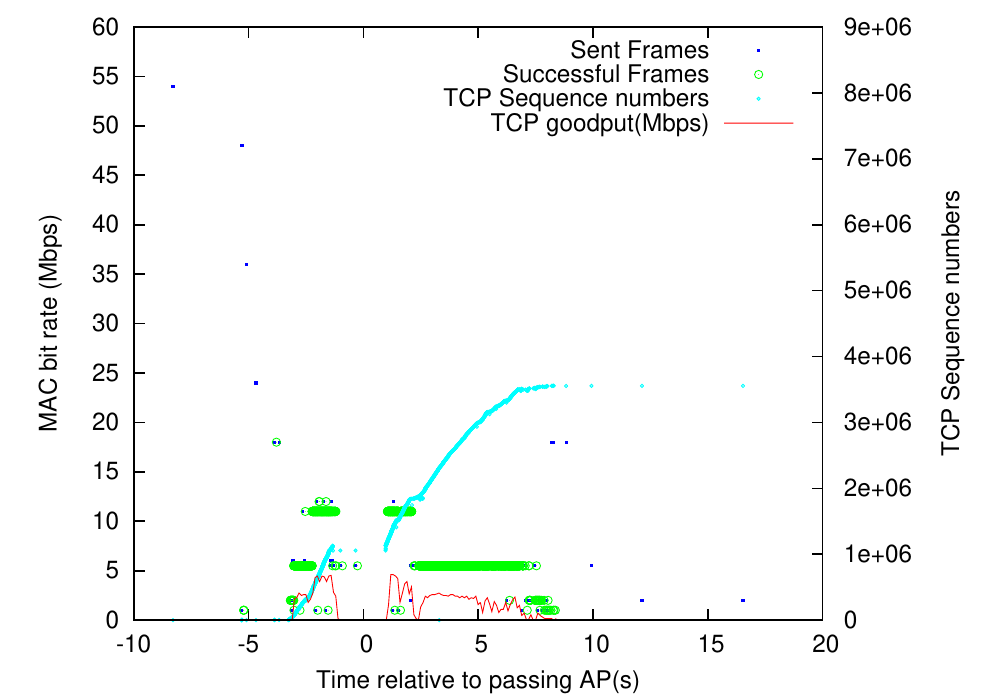 , height=2in, width=3.5in}
\caption{The TCP Sequence numbers and TCP goodput as well as MAC bit rates of AP Frames and Successful Frames against time relative to passing AP for a RRAA run.}
\end{figure}

\begin{figure}[t]
\centering
\epsfig{file=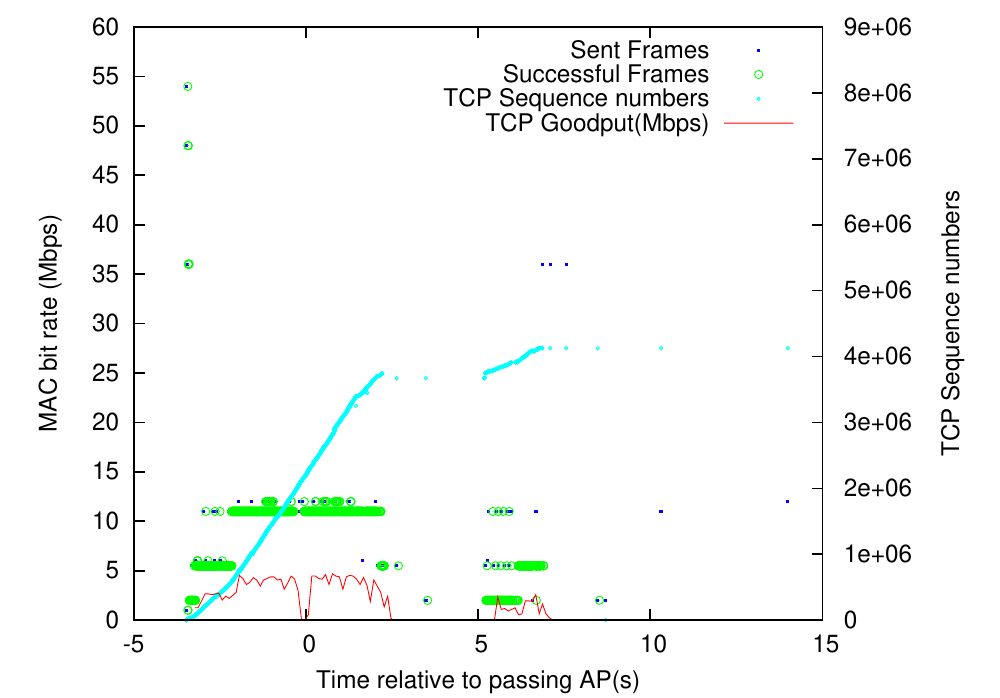 , height=2in, width=3.5in}
\caption{The TCP Sequence numbers and TCP goodput as well as MAC bit rates of AP Frames and Successful Frames against time relative to passing AP for a RRAA run.}
\end{figure}
\noindent \textbf{SampleRate}\newline
SampleRate uses an estimation window of 10s i.e. the statistics of only those packet are considered which were sent in this time window. A large estimation window affects the ability of SampleRate to react to changing channel conditions. However with TCP, a large estimation window was a blessing in disguise for SampleRate. It meant that the rates like 11,5.5,2 and 1Mbps that did well for a longer duration were the ones that were selected most frequently(this can be seen from Figures 25, 26 and 27). Usually only probe packets were sent at other rates. As a result, SampleRate achieved a steady TCP goodput performance. Secondly close to AP they were hardly any TCP losses, resulting in overall better TCP performance as compared to RRAA. The example of SampleRate shows that rate selection algorithms that select rates like 1,2,5.5 and 11Mbps, which have a longer transmission range and lower RSSI threshold, tend to achieve better TCP performance.
\newline \\

\begin{figure}[t]
\centering
\epsfig{file=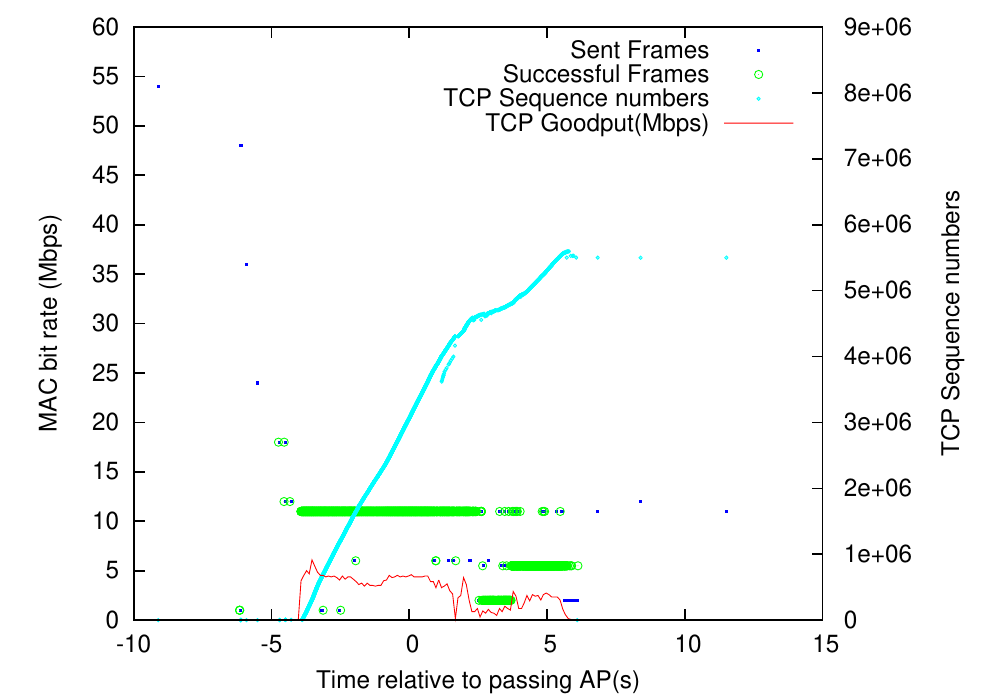 , height=2in, width=3.5in}
\caption{shows the TCP Sequence numbers and TCP goodput as well as MAC bit rates of AP Frames and Successful Frames against time relative to passing AP for a SampleRate run.}
\end{figure}

\begin{figure}[t]
\centering
\epsfig{file=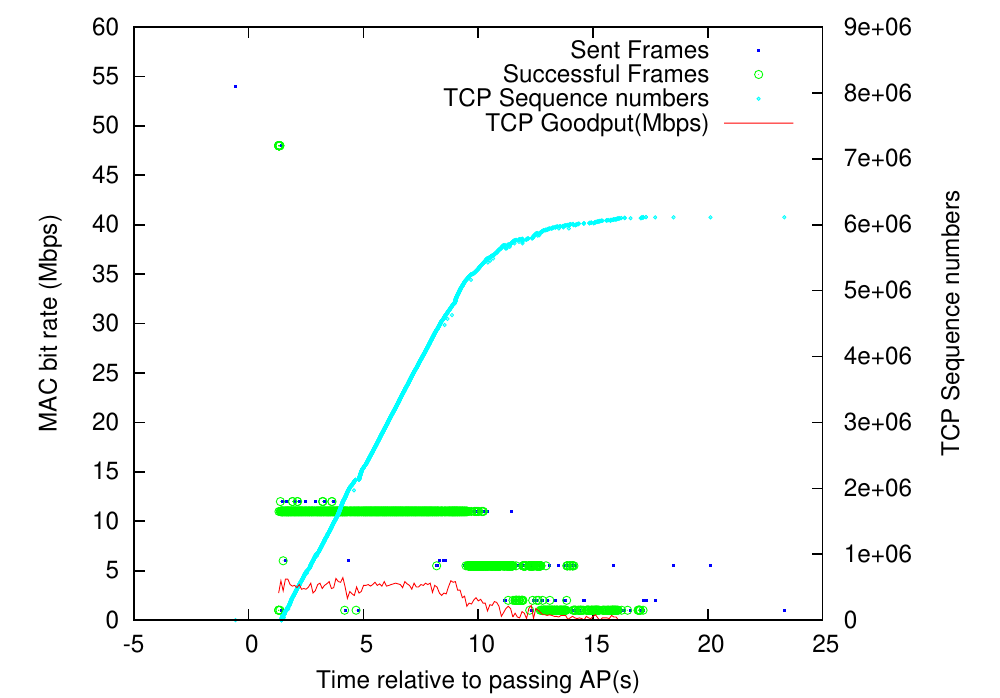 , height=2in, width=3.5in}
\caption{TCP Sequence numbers and TCP goodput as well as MAC bit rates of AP Frames and Successful Frames against time relative to passing AP for a SampleRate run.}
\end{figure}

\begin{figure}[t]
\centering
\epsfig{file=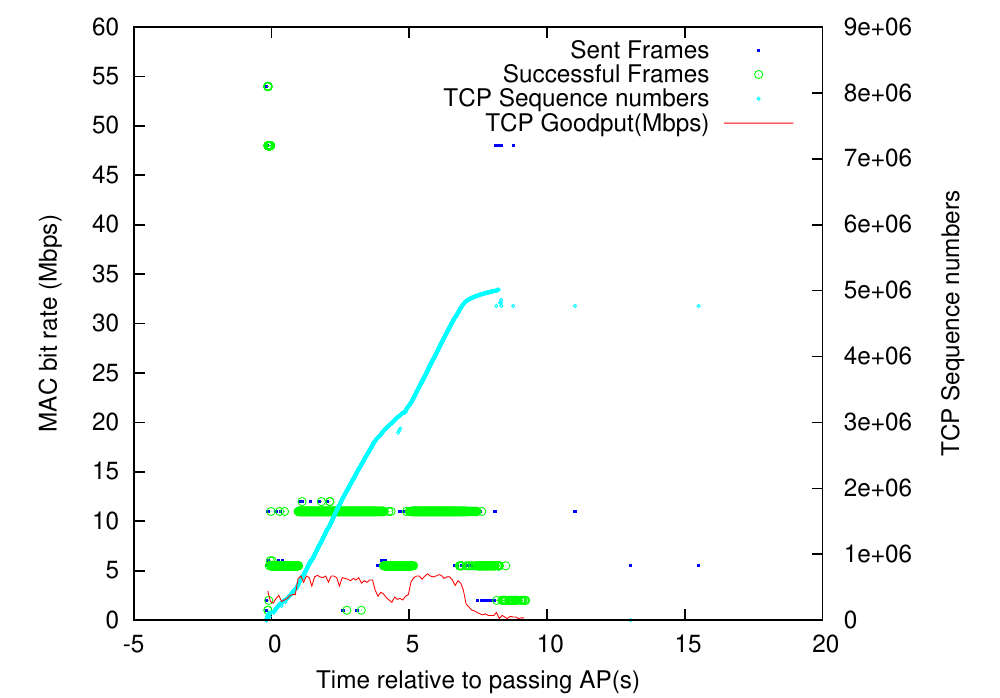 , height=2in, width=3.5in}
\caption{TCP Sequence numbers and TCP goodput as well as MAC bit rates of AP Frames and Successful Frames against time relative to passing AP for a SampleRate run.}
\end{figure}

\begin{figure}[t]
\centering
\epsfig{file=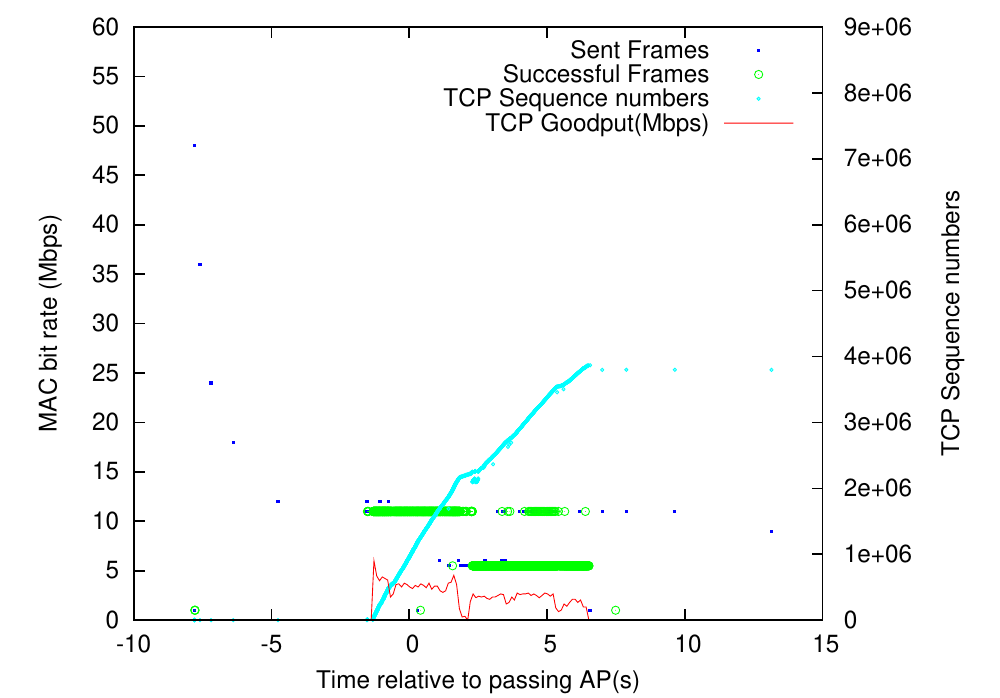 , height=2in, width=3.5in}
\caption{TCP Sequence numbers and TCP goodput as well as MAC bit rates of AP Frames and Successful Frames against time relative to passing AP for a AMRR run.}
\end{figure}

\begin{figure}[t]
\centering
\epsfig{file=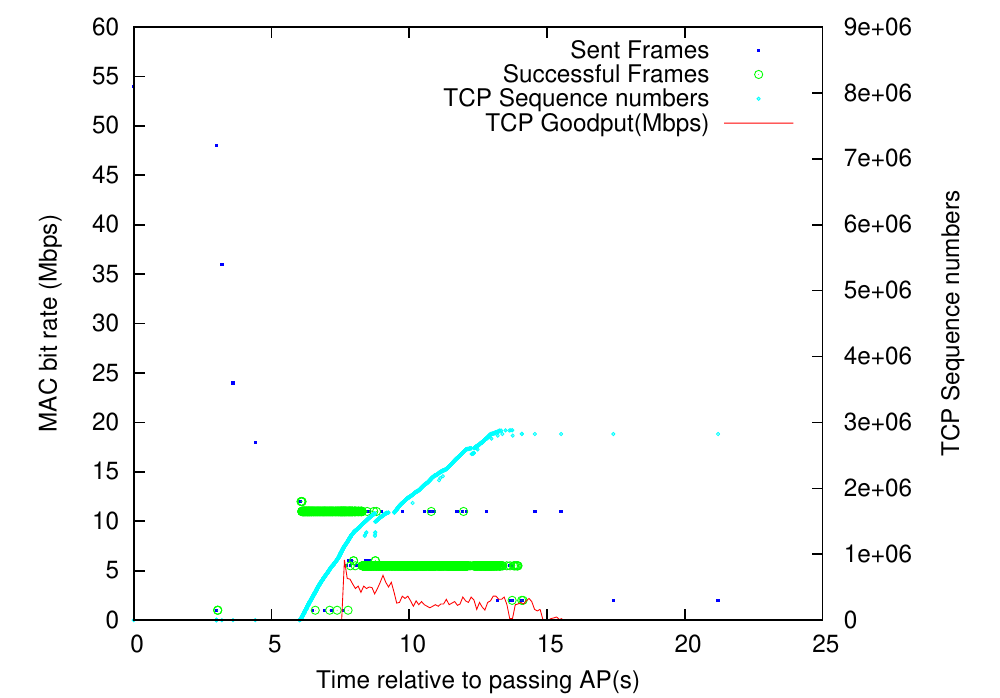 , height=2in, width=3.5in}
\caption{TCP Sequence numbers and TCP goodput as well as MAC bit rates of AP Frames and Successful Frames against time relative to passing AP for a AMRR run.}
\end{figure}

\begin{figure}[h!]
\centering
\epsfig{file=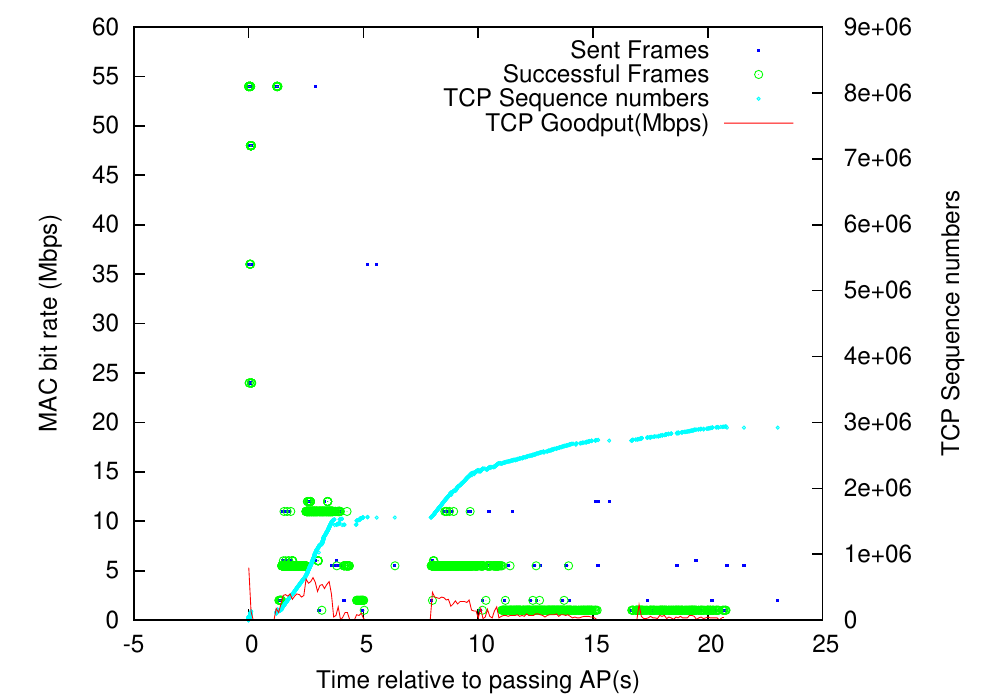, , height=2in, width=3.5in}
\caption{TCP Sequence numbers and TCP goodput as well as MAC bit rates of AP Frames and Successful Frames against time relative to passing AP for a AMRR run.}
\end{figure}

\begin{figure}[h!]
\centering
\epsfig{file=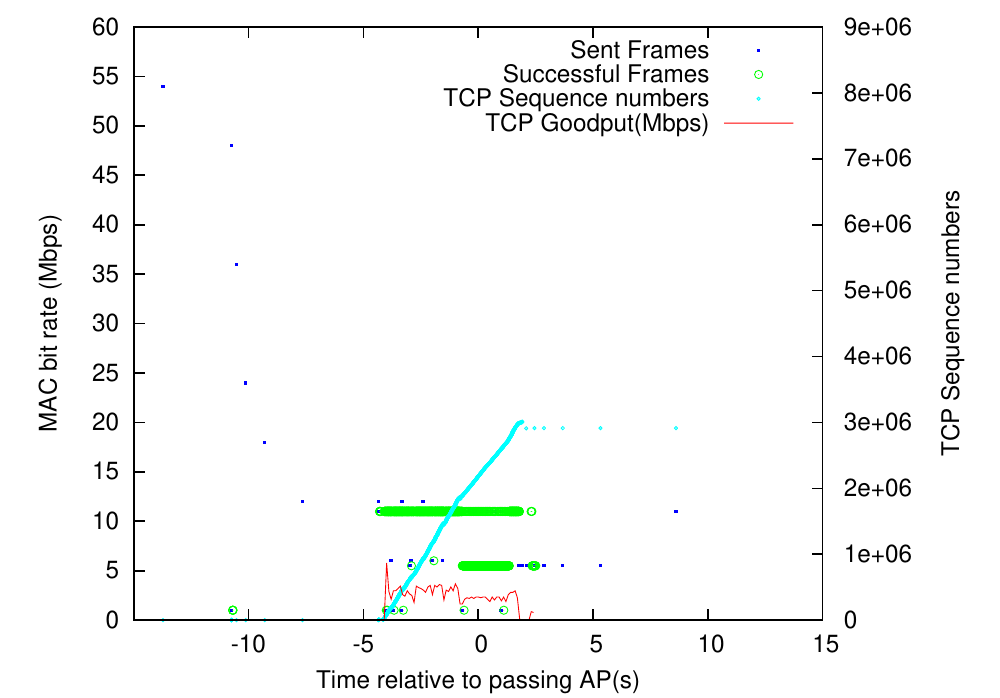 , height=2in, width=3.5in}
\caption{TCP Sequence numbers and TCP goodput as well as MAC bit rates of AP Frames and Successful Frames against time relative to passing AP for a Minstrel run.}
\end{figure}

\begin{figure}[h!]
\centering
\epsfig{file=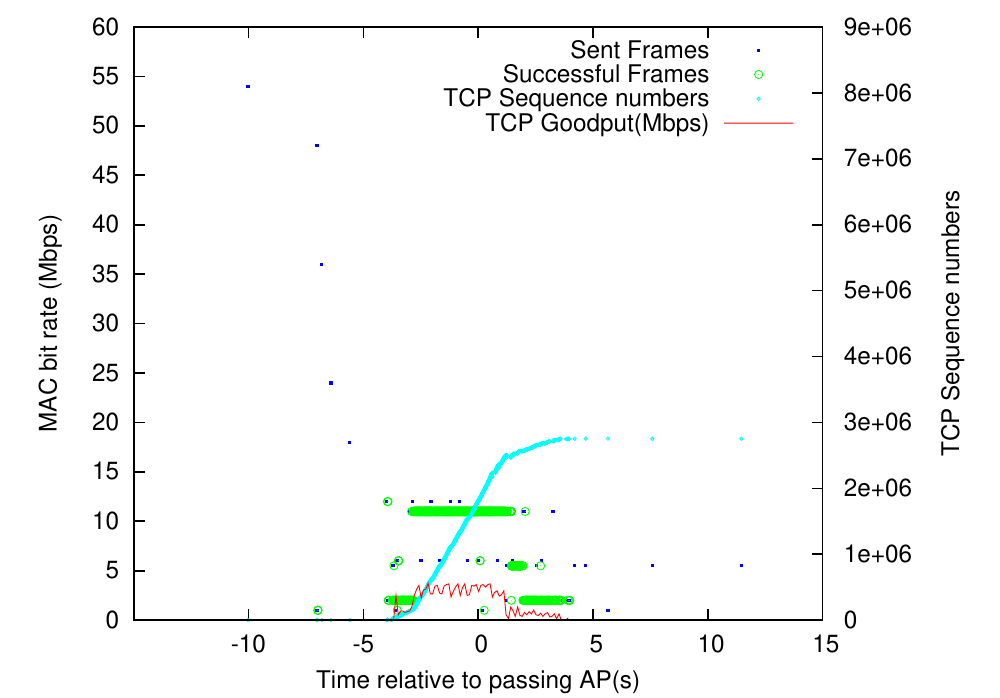 , height=2in, width=3.5in}
\caption{TCP Sequence numbers and TCP goodput as well as MAC bit rates of AP Frames and Successful Frames against time relative to passing AP for a Minstrel run.}
\end{figure}

\begin{figure}[h!]
\centering
\epsfig{file=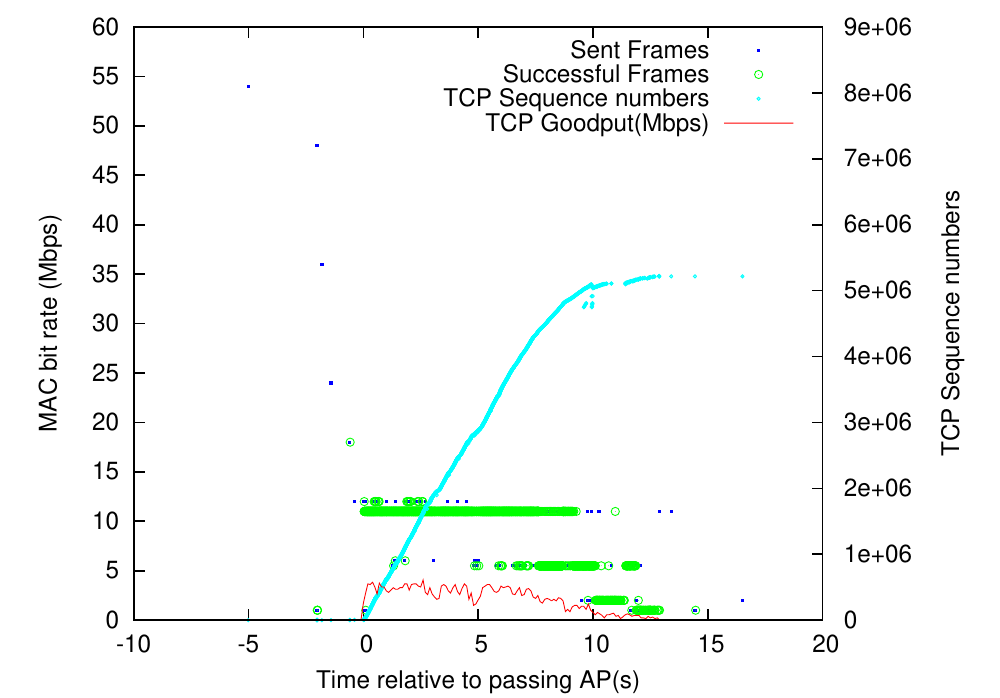, , height=2in, width=3.5in}
\caption{TCP Sequence numbers and TCP goodput as well as MAC bit rates of AP Frames and Successful Frames against time relative to passing AP for a Minstrel run.}
\end{figure}

\noindent \textbf{AMRR}\newline
Like the other two rate adaptation algorithms, AMRR starts off with a high initial MAC bit rate. Specifically, it starts off with a rate of 54Mbps as shown in Figure 28 and 29. It has a time window of 1s. In addition it performs sequential rate switching, hence if a rate is selected which is more than three hops away from the appropriate rate, it will take AMRR more than three seconds to converge to the appropriate rate. That means that in this phase, where AMRR is converging to the suitable MAC bit rate, there will be lots of frame losses, hence there will be TCP timeouts, which will result in TCP invoking exponential backoff. Figure 30 illustrate this point, where in the interval +5 to +10s TCP invokes exponential backoff and the TCP good drops to zero in this period. This again illustrate the point that with TCP, such MAC bit rates should be selected which suffer loss rates to the extent that frame losses are hidden from the TCP. This requires the rate selection algorithms to be aware of the transport layer protocol that they are operating with.
\newline
\newline
\noindent \textbf{Minstrel}\newline
As in the case of the other three rate adaptation algorithms, Minstrel used a high initial MAC bit rate, which often resulted in TCP invoking exponential backoff initially. As a consequence, initially the TCP goodput remained zero for some time.  This is shown in Figures 31, 32 and 33.\newline
Secondly Minstrel selected OFDM rates like 12Mbps and 6Mbps when the channel conditions did not support these rate, resulting in a drop in TCP goodput as shown in Figures 31, 32 and 33. Thirdly 6Mbps was selected when 11Mbps would have performed better as shown by our fixed rate analysis. However EWMA of 75\% meant that Minstrel mostly did not try to use high rates which failed initially, which resulted in Minstrel mostly using lower rates, hence avoiding to many frame losses and consequently packet losses.

\vspace{8mm}
\section{Conclusion}
In this work, we consider Internet access in vehicles, in particular, short-lived connections to roadside 802.11 access points that arise opportunistically as vehicles are in motion. We conduct real outdoor experiments, to investigate the performance of different rate adaptation algorithms, their interaction with higher layers and their impact on the overall connection performance. Specifically, we test four rate  adaptation algorithms namely RRAA, SampleRate, AMRR, and Minstrel, along with TCP bulk traffic and CBR traffic over UDP.  Our experimental results reveal that all the four rate adaptation algorithms used high initial MAC bit rates (e.g., 54 Mbps), often are too slow to adapt to changing channel conditions in vehicular settings (using either a bit rate that is too high or too low), do not take into account the different characteristics of DSSS and OFDM rates and can frequently cause TCP retransmissions.



\newpage
\begin{thebibliography}{10}

\bibitem{iperf}
{Iperf. \emph{http://dast.nlanr.net/Projects/Iperf/}}.

\bibitem{minstrel}
{Minstrel. \emph{http://linuxwireless.org/en/developers/}
  \emph{Documentation/mac80211/RateControl/minstrel}}.

\bibitem{madwifi}
{Multiband Atheros Driver for WIFI. \emph{http://www.madwifi.org/}}.

\bibitem{bicket}
J.~Bicket.
\newblock {Bit-Rate Selection in Wireless Networks. Master's thesis,
  Massachusetts Institute of Technology}.
\newblock February 2005.

\bibitem{prateek}
J.~Bicket.
\newblock {Implementation and Experimental Study of Rate Adaptation Algorithms
  in IEEE 802.11 Wireless Networks. Master's thesis, Iowa State University}.
\newblock 2009.

\bibitem{bychkovsky}
V.~Bychkovsky, B.~Hull, A.~Miu, H.~Balakrishnan, and S.~Madden.
\newblock {A Measurement Study of Vehicular Internet Access Using In Situ Wi-Fi
  Networks}.
\newblock In {\em MobiCom Conference Proceedings}, September 2006.

\bibitem{chen}
X.~Chen, P.~Gangwal, and D.~Qiao.
\newblock {Practical Rate Adaptation in Mobile environments}.
\newblock In {\em {IEEE International Conference on Pervasive Computing and
  Communications(PERCOM)}}, 2009.

\bibitem{cottingham}
D.N.Cottingham, I.~Wassell, and R.~Harle.
\newblock {Performance of IEEE 802.11a in Vehicular Contexts}.
\newblock In {\em {IEEE Vehicular Technology Conference (VTC)}}, 2007.

\bibitem{gass}
R.~Gass, J.~Scott, and C.~D. Thomson.
\newblock {Measurements of In-Motion 802.11 Networking}.
\newblock In {\em IEEE Workshop on Mobile Computing System and Applications
  (HOTMOBILE)}, April 2006.

\bibitem{Giustiniano}
D.~Giustiniano, G.~Bianchi, L.~Scalia, and I.~Tinnirello.
\newblock {An Explanation for unexpected 802.11 Link Level Measurement
  Results}.
\newblock In {\em {IEEE Conference on Computer Communications (INFOCOM)}},
  2008.

\bibitem{hadler}
D.~Hadaller, S.~Keshav, T.~Brecht, and S.~Agarwal.
\newblock {Vehicular Opportunistic Communication Under the Microscope}.
\newblock In {\em MobiSys Conference Proceedings}, June 2007.

\bibitem{camp}
J.Camp and E.Knightly.
\newblock {Modulation Rate Adaptation in Urban and Vehicular Environments:
  Cross-layer Implementation and Experimental Evaluation}.
\newblock In {\em {MobiCom Conference Proceedings}}, 2008.

\bibitem{lacage}
M.~Lacage, M.~Hossein, and T.~Turletti.
\newblock {IEEE 802.11 Rate Adaptation: A Practical Approach}.
\newblock In {\em IEEE MSWiM}, October 2004.

\bibitem{vutukurus}
M.Vutukuru and H.~K.Jamieson.
\newblock {Cross-layer Wireless Bit Rate Adaptation}.
\newblock In {\em {SIGCOMM '09: Proceedings of the ACM SIGCOMM 2009 conference
  on Data communication}}, 2009.

\bibitem{ott}
J.~Ott and D.~Kutscher.
\newblock {Drive-thru Internet: IEEE 802.11b for Automobile Users}.
\newblock In {\em {IEEE INFOCOM}}, 2004.

\bibitem{ott1}
J.~Ott and D.~Kutscher.
\newblock {A Disconnection-Tolerant Transport for Drive-thru Internet
  Environments}.
\newblock In {\em {IEEE Conference on Computer Communications (INFOCOM)}},
  2005.

\bibitem{ramachandran}
K.~Ramachandran, R.~Kokku, H.~Zhang, and M.~Gruteser.
\newblock {Synchronous Twophase Rate and Power Control in 802.11 WLANs}.
\newblock In {\em {ACM Conference on Mobile Systems, Applications, and Services
  (MobiSys)}}, 2008.

\bibitem{shankar}
P.~Shankar, T.~Nadeem, J.~Rosca, and L.~Iftode.
\newblock {CARS: Context Aware Rate Selection for Vehicular Networks}.
\newblock In {\em {IEEE ICNP}}, 2008.

\bibitem{wong}
S.~H. Wong, H.~Yang, S.~Lu, and V.~Bhargavan.
\newblock {Robust Rate Adaptation for 802.11 Wireless Networks}.
\newblock In {\em MobiCom Conference Proceedings}, 2006.

\end{thebibliography}
\end{document}